\documentclass[12pt,a4paper]{article}
\pdfoutput=1

\usepackage[DIV=10]{typearea}

\usepackage[english]{babel}

\usepackage{mathpazo} 

\usepackage[T1]{fontenc}
\usepackage[utf8]{inputenc}

\PassOptionsToPackage{pdftex,hyperfootnotes=false,bookmarks=false,pdfpagelabels}{hyperref}
\usepackage{hyperref}  
 \usepackage{tcolorbox}

\definecolor{DarkBlue}{rgb}{0,0,0.5}
\definecolor{NiceGreen}{RGB}{0,153,72}
\hypersetup{%
    colorlinks=true, linktocpage=true, pdfstartpage=3, pdfstartview=FitV,%
    breaklinks=true, pdfpagemode=UseNone, pageanchor=true, pdfpagemode=UseOutlines,%
    plainpages=false, bookmarksnumbered, bookmarksopen=true, bookmarksopenlevel=1,%
    hypertexnames=true, pdfhighlight=/O,
    urlcolor=webbrown, linkcolor=DarkBlue, citecolor=NiceGreen, 
}

\usepackage{amsmath}
\usepackage{bbm}
\usepackage{amsfonts}
\usepackage{amssymb}
\usepackage{amsthm}

\usepackage{fancyhdr}

\usepackage{enumerate}

\usepackage{multirow} 
\usepackage{float}
\usepackage{subcaption}
\usepackage{graphicx}
\usepackage{epsfig,epstopdf}
\usepackage{xcolor}

\usepackage{pgfplots}
\usetikzlibrary{pgfplots.groupplots}
\pgfplotsset{compat=1.11}
\usepgfplotslibrary{fillbetween}

\usepackage{tikz}
\usetikzlibrary{
  shapes,
  shapes.geometric,
  positioning,
  fit,
  fadings,
    pgfplots.groupplots,
  }

\usepackage{braket}
\usepackage{color}
\usepackage[round,sort,comma]{natbib}

\usepackage{setspace}   
\makeatletter 
\hypersetup{pdftitle = {Analogue Quantum Simulation: A Philosophical Prospectus},
	     pdfauthor = {Dominik Hangleiter, Jacques Carolan, Karim Thebault},
	     pdfsubject = {philosophy of science, quantum physics},
	     pdfkeywords = {Quantum simulation,
	     	quantum information,
	     	understanding,
	     	explanation
		    certification, 
		    computer simulation,
		    analogical inference,
		    quantum computation,
		    quantum speedup
		    }
	    }
\makeatother

\title{Analogue Quantum Simulation: \\ A Philosophical Prospectus}

\author{Dominik Hangleiter\footnote{Fachbereich Physik, Freie Universit\"at Berlin, 14195 Berlin, Germany.} , 
Jacques Carolan\footnote{
Department of Electrical Engineering and Computer Science, Massachusetts Institute of Technology, Cambridge, Massachusetts 02139, United States.} , 
and Karim Th\'{e}bault\footnote{Department of Philosophy University of Bristol, Bristol, United Kingdom.}}

\begin{document}

\maketitle

 \begin{abstract}
 
This paper provides the first systematic philosophical analysis of an increasingly important part of modern scientific practice: analogue quantum simulation. We  introduce the distinction between `simulation' and `emulation' as applied in the context of two case studies. Based upon this distinction, and building upon ideas from the recent philosophical literature on scientific understanding, we provide a  normative framework to isolate and support the goals of scientists undertaking analogue quantum simulation and emulation. We expect our framework to be useful to both working scientists and philosophers of science interested in cutting-edge scientific practice.

 \end{abstract}

\tableofcontents

\section{A New Instrument of Science}

An analogue quantum simulator is a bespoke device used to simulate aspects of the dynamics of another physical system using continuous parameters. An important example of an analogue quantum simulator is provided by a `source system' comprised of ultracold atoms confined to an optical lattice. In a certain regime this system has been found to realize Hubbard Hamiltonians that describe `target systems' that exhibit strongly correlated many body physics. Whereas classical analogue simulators have largely been superseded by classical computer simulation, quantum analogue simulators provide the most plausible near-term device for efficiently simulating a wide range of quantum systems. In particular, with increasing awareness of the immense difficulties in constructing a universal quantum computer, analogue quantum simulation has emerged as a critical technique towards the near-term understanding of quantum systems. For models that are computationally intractable by conventional means, or for systems which are inaccessible or experimentally challenging to manipulate, analogue quantum
simulation provides a uniquely powerful new inferential tool.  

It is an open question whether analogue quantum simulation is a new mode of
scientific inference, or whether it reduces to more traditional modes of
inference such as analogical argument, computer simulation, or experimentation.
What is more, it is even unclear whether analogue simulation refers to one and
the same activity across the very different fields in which the term
appears.\footnote{For sake of both brevity and clarity, in this paper we will
restrict ourselves to the analysis of analogue quantum simulation of
non-gravitational systems. That is, we will not touch upon analogue simulation
as found in the increasingly exciting field of analogue gravity. See
\citep{Dardashti:2015} and \citep{thebault:2016}. Comparison between the two species of analogue simulation (i.e. gravitational and non-gravitational) is an important challenge for future work.} One principal aim of this paper is to situate analogue quantum simulation on the `methodological map' of modern science \citep{galison1996} and in doing so clarify the functions that analogue quantum simulation serves in scientific practice. To this end, we will introduce a number of important terminological distinctions, most importantly that between `simulation'  and `emulation'. We will then consider two case studies: ultracold atomic simulation of the Higgs mode in two dimensions and photonic emulation of quantum effects in biological systems. Building upon the discussion of \cite{ToyModels}, we will then isolate three key goals of scientists undertaking quantum simulation and emulation, based upon the distinction between the (modally stronger) notion of `how-actually understanding' and the (modally weaker) `how-possibly understanding'. The comparison with more traditional forms of scientific inference will then inform our placement of analogue simulation and emulation on a methodological map based upon forms of understanding. 

Building on the foregoing more interpretive work, we then provide a normative analysis of analogue simulation and emulation. We will propose that the types of validation and certification that are appropriate to a particular case of simulation or emulation can be identified by considering the form(s) of understanding the scientists are aiming to acquire. In this sense, we will provide epistemic norms for the practice of analogue simulation and emulation. It also seems reasonable to require further pragmatic norms in the application of quantum simulation or emulation. Such norms pertain to issues around the usefulness of a simulation towards goals that are not directly epistemic. The final section will offer two suggestions in this regard: a heuristic norm of `observability' and a computational norm of `speedup'. Together we expect that our epistemic and pragmatic norms will support the successful practice of analogue quantum simulation and emulation in both present day and future science.

\section{Distinctions with a Difference}

The term `analogue quantum simulation' is becoming widely popular in the context of quantum information science and carries with it a number of subtle and interrelated connotations. In order to better understand the contemporary scientific practice relating to analogue simulation it is important to make a number of terminological and methodological distinctions:
\begin{itemize}
\item [D1.] \textbf{Analogue vs.\ Analog}. The first distinction is between the two senses `analogue' that are at play. The first sense relates to the idea that a physical system is `comparable to' another physical system in some relevant sense. This sense of analogue brings to mind the idea of an `analogue model' or `argument by analogy' and in this vein there is an interesting connection with relevant philosophy of science literature of the last half century. We will return to this connection in \S \ref{interpretive}. 

\item [D2.] \textbf{Analog vs.\ Digital} The second sense of analogue in analogue simulation relates to the idea that the simulation in question is `not digital'.\footnote{In the paragraphs that follow we will consistently use the British spelling analogue for the `similar to' sense and American spelling analog for the `not digital' sense.}
  While in a digital simulation, the computational model is based on a discretization of both the input and output encoding of the computation as well as the control parameters of the computation itself, in an `analog' simulation the computation remains continuous throughout. In particular, this is often the case for the time-parameter in a simulation of the dynamical evolution of a physical system.  Typical analog signals include currents, optical power or turns of a gear.\footnote{It is important to note here that we are distinguishing here between analog vs digital devices rather than representations. See \cite{goodman:1968,lewis:1971,trenholme:1994,maley:2011} for discussion of the latter distinction.} 

\item[D3.] \textbf{Reprogrammable vs.\ Bespoke} We can further classify analog devices that can be reconfigured to implement distinct tasks as `reprogrammable' analog simulators. An early example of such a reprogrammable analog simulator is the differential analyser built by Bush in 1931 at MIT \citep{sep-computing-history}. In contrast, we have `bespoke' analog simulators designed to implement a specific task (or small sets of tasks). The same distinction is possible, though increasingly unusual, in the case of digital simulators. That is, we can have digital simulators that are built to implement a specific task such as running a particular chess algorithm.

\item[D4.] \textbf{Classical vs.\ Quantum}. Next we can distinguish (at least in practice) between simulators that are built out of classical components from those built out of quantum components. In our terminology, a desktop computer is a reprogrammable classical digital simulator -- it is classical since it implements programs using classical logic gates. We can contrast this with a reprogrammable quantum digital simulator like IBM's 20-qubit quantum computer\footnote{See  https://researchweb.watson.ibm.com/ibm-q/}-- it is quantum since it encodes information on individual quantum systems and cannot be efficiently simulated classically.\footnote{Here, `efficient' refers to a computational runtime (or space usage) which scales at most polynomially in the size of the problem.}\textsuperscript{,}\footnote{Making a clear in principle distinction between classical and quantum simulation is an extremely subtle problem. See \cite{sep-qt-quantcomp,cuffaro:2017a,cuffaro:2017b} for relevant discussion.} The philosophical literature on classical digital simulation contains a number of signifiant points of analysis and controversy that will be relevant for our analysis. Again we will return to this connection in \S \ref{interpretive}. 
\end{itemize}

All four of these distinctions are philosophically interesting and could be subject to a lengthy further discussion independent of each other. However, what is crucial for our purposes is that there is a clear demarcation between simulation via a reprogrammable digital quantum device and simulation via a bespoke analog quantum device. The digital quantum device in question is constructed out of quantum logic gates and, given that the algorithm implemented by these gates can be changed by reordering their connections, constitutes a programmable `quantum computer'. A quantum computer that comprises a universal gate set can efficiently implement any quantum algorithm and is called a `universal quantum computer'.  It can be proven that such a machine would provide us with the capacity to perform a digital simulation of any physical quantum system that features local interactions \citep{lloyd_universal_1996}.
Intrinsic to the nature of quantum computation is that coherent quantum states are extremely fragile.
However, one of major breakthroughs of the field came with the advent of error correction, meaning that provided the error incurred during a computation is below some fault-tolerance threshold, arbitrarily long error-free computations may be performed \citep{Devitt:2013hb}.
The experimentalists task is thus to engineer such systems with extremely high fidelity and on a sufficiently large-scale: a monumental challenge.

With increasing awareness of the immense difficulties in constructing a large-scale fault-tolerant quantum computer, significant appetite has emerged for small-scale non-universal quantum devices that can solve specific tasks.
In particular, an analogue quantum simulator is not constructed out of quantum logic gates. Rather, it is a well controlled quantum system which can be continuously manipulated to implement dynamics of interest: namely, that of a system (be it an abstract model or a concrete physical system) which the experimentalist does not have direct epistemic access to.
The continuous nature of the computation means that there is no known way to error correct these systems, which in turn makes scaling arguments problematic \citep{Hauke:2012dq}.
However, it is hoped there exist simulation regimes wherein error effects are not overwhelming yet classical computers fail \citep{Aaronson:2010tja,bremner_average-case_2016}, or that errors in the simulator properly correspond to errors in the physical system being simulated \citep{Cubitt:2017ti}.  

The final and crucial terminological distinction is between two different implementations of analogue quantum simulation:
\begin{itemize}
\item [D5.] \textbf{Simulation vs.\ Emulation}. In an analogue quantum `simulation', features of the source system we are manipulating are being appealed to for the specific purpose of gaining knowledge pertaining to features of an abstract theoretical model. In contrast, in an analogue quantum `emulation', features of the source system we are manipulating are being appealed to for the specific purpose of gaining knowledge directly pertaining to features of an actual and concrete physical system. 
\end{itemize}
It should be noted here that the terminology we are using is our own and thus our use of `emulation' need not coincide with use of the term elsewhere. 
Our distinction between simulation and emulation is explicitly routed in the intentions of the scientists performing the simulation or emulation. 
That is, in principle the \textit{same experiment} might be considered a simulation in one context and an emulation in another, merely in virtue of the relevant scientists having differing targets about which they wish to gain knowledge. 
Consequently, it is also possible that a scientist may have a dual purpose in carrying out a given analogue experiment on a source system: they may principally wish to gain knowledge pertaining to features of or phenomena arising in an abstract theoretical model, but have a secondary interest in features of an actual and concrete physical system that this model represents.
In such cases the natural analysis would be that a given experiment has components of both simulation and emulation. 
For the purposes of the remainder of the paper we will make the idealisation that the analogue experiments in question are pure cases of simulation or emulation. 
We will return to these questions of ambiguity in final section. 
In the following section we will present two detailed case studies in order to illustrate the differences that ground the simulation vs.\ emulation distinction further.
We will then, in Sections \ref{interpretive} and \ref{map}, proceed to discuss how simulations and emulations are used to learn about features pertaining to abstract and concrete phenomena respectively.

\section{Case Studies}
\label{descriptive}

In this section we will provide a detailed description of two case studies taken from cutting -edge scientific practice in analogue quantum simulation and analogue quantum emulation. Although we will not shy away from providing the formal details behind the physics of our models, the non-technical reader will be able to glean the most significant scientific details by reading the non-technical summary provided at the start of each case study whilst refering to the associated diagrammatic schema.  

\subsection{Cold Atom Simulation and the Higgs Mode}
\label{higgs cs}

\paragraph{Summary and Schema}

Cold atoms in optical lattices are one of the most important platforms for quantum simulations \citep{Jaksch1998,Greiner2002, OptLatQS}. 
In such systems an artificial lattice potential is created using counterpropagating laser beams: a crystal of light. 
The resulting intensity pattern acts as a space-dependent lattice potential for certain atoms via the dipole-dipole coupling between the light field and the
dipole moment of these atoms. Such optical-lattice potentials can be combined with so-called magneto-optical traps (MOT) using which one can create a low-temperature state of a confined atomic cloud consisting of atoms such as $^{87}$Rb (a bosonic atom) or $^{40}$K (a fermionic one). By adding the optical-lattice potential to the MOT potential one can thus realise a system in which hundreds to thousands of atoms evolve coherently while interacting among themselves and propagating through the lattice. These systems bear strong similarities to real solid-state systems, where one encounters the same lattice structure for the potential of electrons hopping between atoms. Strikingly, the Hamiltonian that has been found to accurately describe the dynamics of cold atoms in optical lattices is the Hubbard Hamiltonian \citep{Jaksch1998}. In its fermionic variant, the Hubbard Hamiltonian is the simplest model describing interacting fermions that features Coulomb repulsion, a nontrivial band structure, and incorporates the Pauli Principle \citep{Hubbard-1963}. While the experimental realisation of the Fermi-Hubbard model remains an outstanding challenge, the bosonic variant is being realised in many laboratories across the world today. What makes cold atoms in optical lattices so well suited for the simulation of condensed-matter systems is the possibility to both manipulate and probe these systems with high precision. 

An example of a quantum simulation that has been performed using cold atoms in optical lattices is the study of the Higgs mechanism in two dimensions \citep{Higgs:2012}. 
The Higgs mechanism appears in the study of spontaneous symmetry breaking,
which lies at the heart of our understanding of various natural phenomena. 
Most famously, the Higgs mechanism appears in particle physics where a spontaneously broken symmetry leads to the emergence of massive particles. The Higgs mechanism is also important in condensed matter physics, where the phases of matter can most often be understood in terms of breaking symmetries. It is a subject of theoretical controversy \citep{sachdev_universal_1999,altman_oscillating_2002,podolsky_visibility_2011,podolsky_spectral_2012,pollet_higgs_2012,liu_massive_2015} whether in two-dimensional systems a Higgs mode is present. Analytical solution or Quantum Monte Carlo methods fail and perturbative methods fail to provide a definite answer. This is precisely the situation where analogue quantum simulation proves a powerful new inferential tool. In a critical superfluid--Mott insulator (SF-MI) transition the equations describing cold atoms in an optical lattice are given by an $O(2)$-symmetric field theory \citep{altman_oscillating_2002}. It is thus possible to probe the solution space of the $O(2)$-symmetric field theory by inducing a SF-MI critical transition in the cold atom system. In one experiment, researchers attempted to answer the theoretical question of whether a Higgs amplitude mode exists in two dimensions, by performing an experiment on cold atoms in optical lattices \citep{Higgs:2012}. The experimental findings (detailed below) suggested the signature of spontaneous symmetry breaking with a two-dimensional Higgs mode. This case study illustrates our notion of analogue quantum simulation since features of the source system (Higgs signature in the ultracold atom system) that is being manipulated are being appealed to for the specific purpose of gaining knowledge pertaining to features of an abstract theoretical model (Higgs mode in a two dimensional field theory). This is not a case of analogue quantum emulation since the intention of the experimenters is not to gain knowledge directly pertaining to features of an actual and concrete physical system.

\begin{figure}
\label{higgs}
\centering
    \includegraphics[height=0.3\textheight]{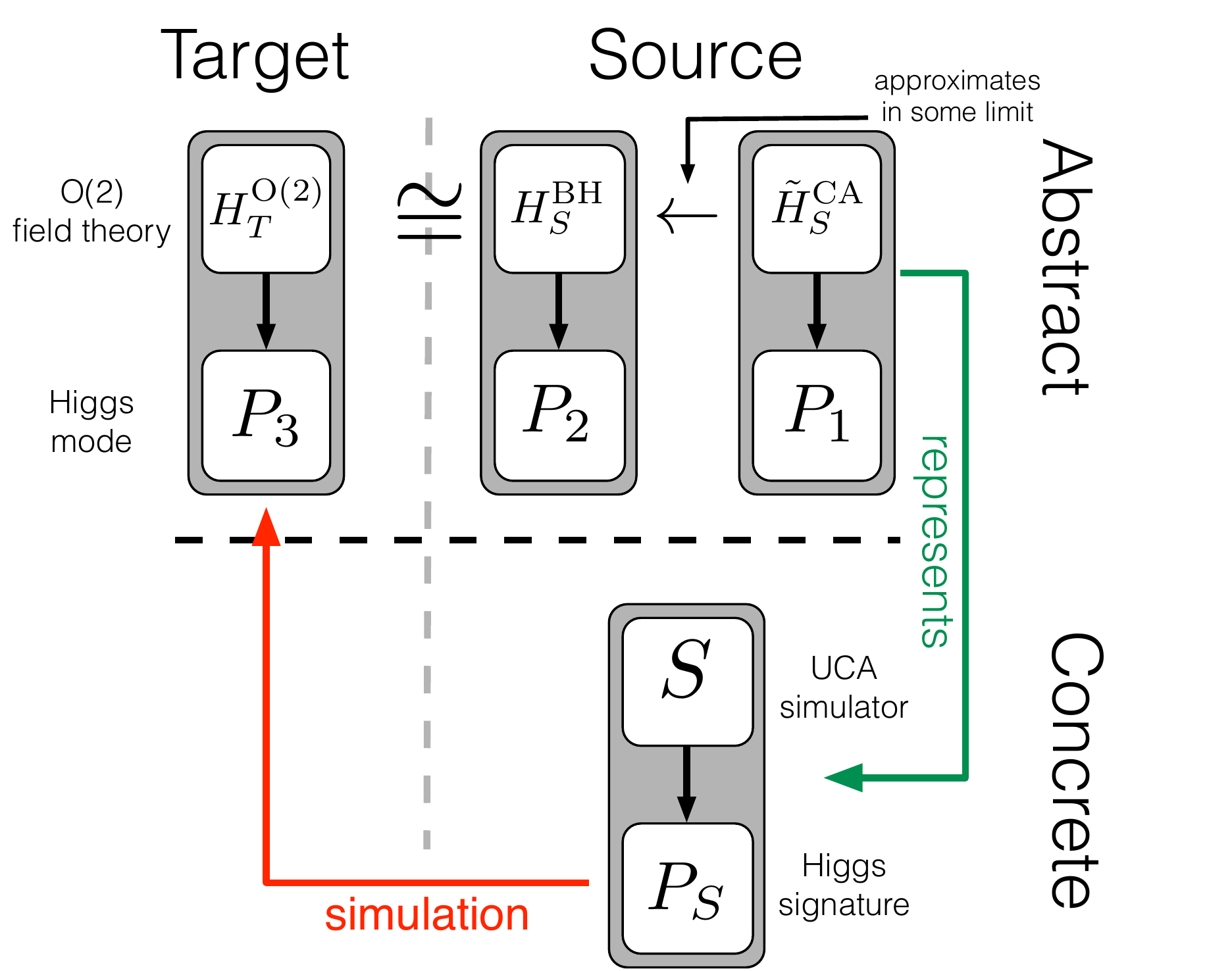}
\caption{Schema for Analogue Simulation Case Study (see main text for figure explanation).
    \label{fig: higgs simulation schema} 
    }
\end{figure}

Fig.~\ref{fig: higgs simulation schema} provides a schema for the general structure of inferences in analogue quantum simulation based upon our case study.\footnote{Our diagrammatic language is inspired by the abstract framework for discussing analogue simulation introduced by \cite{Dardashti:2015}. This in turn built on earlier work by \cite{Winsberg-2010}.} The true Hamiltonian that represents a cold atom system, $\tilde{H}_S^{CA}$, is approximated in a certain parameter regime by the Bose-Hubbard Hamiltonian, $\tilde{H}_S^{BH}$. In the vicinity of the critical point the long-wavelength, low-energy dynamics given by $\tilde{H}_S^{BH}$, then corresponds to that of an an $O(2)$-symmetric field theory with Hamiltonian $\tilde{H}_T^{O(2)}$. $\tilde{H}_S^{CA}$ stands in a representation relation\footnote{We note that `representation' is a heavily loaded word within the philosophy of science. Although our approach here is to be as neutral as possible with regard to characterisations of representation, for our purposes the relevant sense of representation here and elsewhere can be read as simple denotation. See \cite{frigg:2017} for an interesting discussion of a more sophisticated sense of representation that may or may not be relevant here.} with the ultracold atom source system, $S$. Analogue quantum simulation is then a relation between $S$ and $\tilde{H}_T^{O(2)}$. Analogue quantum simulation is a relation between a concrete source system and an abstract target model. Crucially, there is no concrete target system and so the bottom left hand side is empty. The goal of analogue quantum simulation is not to gain understanding of actual phenomena in a concrete physical system. The importance of this feature for the distinction between simulation and emulation will become clear in the second case study, considered in \S \ref{emulation cs}.

\paragraph{Bose-Hubbard Physics} 

As already noted, quantum simulation of Hubbard physics is significant part of contemporary scientific practice. The (abstract) phenomena that may be studied in the context of Hubbard physics include the non-equilibrium quantum dynamics leading to an equilibrated or thermalised state \citep{BlochEisertRelaxation,Schreiber2015mbl,Bloch2016mbl}, 
quantum phase transitions \citep{Greiner2002, Braun-pnas-2015,
esslinger2016phases}, 
magnetism \citep{Struck2011magnetism,murmann2015heisenberg}, metal-insulator
transitions, and high-temperature superconductivity \citep{ColdFermions}. 
The Bose-Hubbard Hamiltonian is given by \citep{Jaksch1998} 
\begin{align} 
  \label{bose hubbard}
  H_{BH} = - J \sum_{\langle j,k \rangle} b_j^\dagger b_k + b_j^\dagger b_k + \frac{U}{2}
  \sum_j b_j^\dagger b_j^\dagger b_j b_j \, . 
\end{align}
Here, $b_j^{\dagger}$ denotes a bosonic annihilation (creation) operator at
site $j$, $U$ denotes the energy cost from having two atoms on the same site,
while $J$ is the energy gain when hopping from one site to the next. 
Using the response of the atoms to an external magnetic field as well as amplitude and phase of the generating laser beams one can tune both $J$ and $U$ independently. What is more, by superimposing several lattice potentials with different wavelengths one can
even realise next-nearest-neighbour interactions and lattices with higher
periodicity \citep{Foelling-Nature-2007}. 
It is even possible to realise low (one and two) dimensional systems by increasing the potential barriers in the orthogonal directions to suppress tunnelling.
In such a way one is able to access a large parameter regime and probe a variety of phenomena that occur in the many-body system. 

There is a large variety of methods available to probe the system which stands in stark contrast to real solid state systems. 
Using highly
focused microscopes one can probe the lattice on the single-atom level
`in vivo' \citep{Bakr-Science-2010, Sherson-Nature-2010}. 
Using so-called time-of-flight imaging and
variants thereof one can also measure the free-space and quasi-momentum
distribution of the atoms \citep{Foelling-Nature-2007,Bloch-RMP-2008}. Thus, using cold atoms in optical lattices one is able to simulate the
physics of complex condensed-matter systems in a well-controllable and accessible many-body 
system.  
Indeed neither the bosonic nor the fermionic variant of the Hubbard Hamiltonian
admits an analytical solution or is amenable to numerical simulations outside
of certain parameter regimes \citep{BlochEisertRelaxation}.

\begin{figure}
\centering
    \includegraphics[width = .9\textwidth]{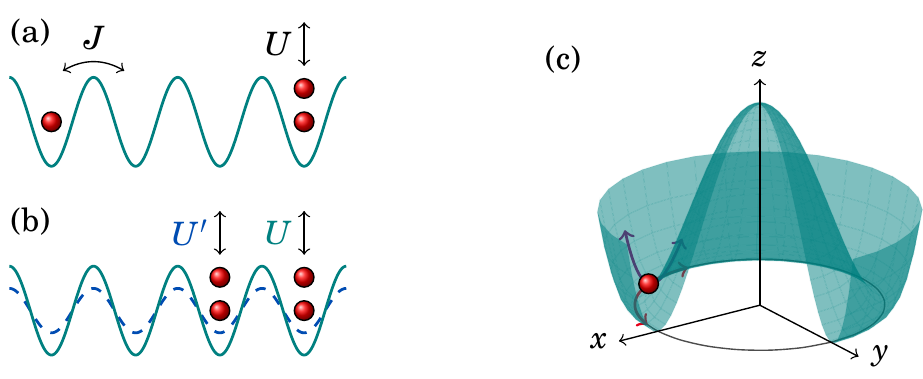}
    \caption{
    (a) Illustration of the experimental system. The atoms are confined by a lattice potential subject to the Bose-Hubbard Hamiltonian \eqref{bose hubbard} with hopping strength $J$ and on-site interaction $U$ . 
    (b) In the experiment of \citep{Higgs:2012} the lattice depth is modulated
    by 3\% whereby the interaction constant $U$ is modulated similarly. 
    (c) Illustration of the Higgs mode (purple arrow) and Nambu-Goldstone modes
    (red arrow) in a Mexican Hat potential. 
        \label{higgs ol}}
\end{figure}

\paragraph{Broken symmetries in $O(2)$-symmetric field theory.} 

Important examples of broken symmetries are the emergence of magnetic or
superconducting states of matter below a certain temperature $T_c$.
For example, iron is paramagnetic above $T_c$ and thus only responds to
external magnetic fields. However, below $T_c$ it is ferromagnetic and thus remains magnetised even in the absence of external magnetic fields. 
In this case the rotational symmetry of the elementary magnets present in the metal is broken as the
temperature sinks from above $T_c$ to below $T_c$; above $T_c$ there is no
preferred direction, while below $T_c$ there is a preferred axis. 
Thus the rotational symmetry described by the symmetry group $O(3)$ is broken. 
A witness for a broken symmetry is called an order parameter. This is a quantity that turns nonzero only as the respective symmetry is broken. 
For example, in the case of the broken rotational $O(3)$ symmetry of the elementary
magnets in iron, an order parameter is given by the total magnetisation of the system,
a quantity that is zero in the paramagnetic phase and nonzero in the ferromagnetic phase. 

Consider the case where we would like to rotate the direction of
magnetisation by an angle. 
The energy cost of a global rotation of the magnetisation is zero as the
relative orientation of the spins does not change at all. 
However, such a global rotation is unlikely to take place as all spins would need to `coordinate'. 
In contrast, infinitesimal rotations of individual spins have an infinitesimally small energy cost. 
As an infinitesimal rotation of a single spin is excited there are two infinitesimal-energy excitations (called Goldstone modes) trying to restore the original direction of magnetisation as dictated by the remaining (aligned) spins.
This is because for an $N$-dimensional order parameter, there are $N-1$ directions in which rotations are possible. 
On the other hand, we can also try and excite an amplitude change of the magnetisation. 
An amplitude excitation, in turn, requires a finite excitation energy. 
This results in a single so-called Higgs mode. 
Famously, in particle physics the Higgs mechanism explains the emergence of
particles' masses in terms of the broken symmetry. 
However, it also features in the explanation of superconductivity and superfluidity as a fundamental collective excitation. 

Let us make this intuitive picture more precise using a simple model for Higgs and Goldstone modes is described by a
$O(2)$-symmetric complex field $\Psi = |\Psi| \mathrm{e}^{\mathrm{i} \phi}$ that is governed by a potential with a characteristic Mexican-hat shape (Fig.~\ref{higgs ol}) in the ordered phase \citep{SchollwoeckASP}. The order parameter for this phase $A = |\Psi| $ takes a non-zero value in the minimum of this
potential and the phase acquires a definite value via spontaneous symmetry breaking of the rotational symmetry on the circle. 
We can now expand the field around this symmetry broken ground state and find
two elementary excitations: i) a transversal (Goldstone) mode along the
minimum of the potential; and ii) a longitudinal amplitude (Higgs) mode along the radial direction. While the Higgs mode requires a finite excitation energy and is therefore
gapped (has a mass), the Goldstone mode is gapless (and therefore massless). 
These correspond precisely to the excitations possible for the case of the elementary magnets discussed above. 
As the disordered phase is approached, the central peak of the hat drops down
to zero which leads to a characteristic `softening' of the finite excitation energy (the gap) of this mode.

The `smoking gun' feature of the Higgs mode is a resonance-like feature in the scalar
susceptibility, that is, the correlation function of $A^2 = |\Psi|^2$ \citep{huber_dynamical_2007,huber_amplitude_2008,podolsky_spectral_2012}. 
This quantity is proportional to the energy absorbed by the system in response 
to external driving as a function of the driving frequency $\nu$ and
can be measured via the temperature change of the system \citep{liu_massive_2015}. 
It has been a subject of controversy
\citep{sachdev_universal_1999,altman_oscillating_2002,podolsky_visibility_2011,podolsky_spectral_2012,pollet_higgs_2012}
whether in two-dimensional systems such a Higgs mode is present, 
or whether it becomes overdamped via coupling to Goldstone modes
resulting in a low-frequency divergence \citep{Higgs:2012}. 
This is
due to the fact that even the simplest relativistic field theory as described
above remains elusive to analytical or numerical treatment. 
Both analytical solution and Quantum Monte Carlo methods
fail. Hence, approximations in the form of mean-field theory or perturbative treatment are necessary in the vicinity of the quantum phase transition and only certain parameter regimes are accessible. 

\paragraph{The Higgs mode in the superfluid--Mott insulator (SF-MI) transition} 

In order to clarify the question whether a Higgs amplitude mode persists in two dimensions, \citet{Higgs:2012} performed an experiment using ultracold atoms in optical lattices. 
At unit filling and zero temperature this system undergoes a quantum phase transition between a
superfluid (ordered) and a Mott insulating (disordered) phase at a critical value $j_c$ of $j = J/U$. 
While in the Mott phase the atoms are strongly localized to the minima of the
potential wells, in the superfluid phase they are delocalized across the entire
lattice. This phase transition is effectively described by a relativistic
$O(2)$-symmetric quantum field theory as explained above with order parameter $\Psi(x_i) = \sqrt{\overline{n}} \langle b_i \rangle$, where $x_i$ is the
position of site $i$ and $\overline{n}$ is the mean atom density in the gas \citep{SchollwoeckASP,altman_oscillating_2002}. 
The expectation value $\langle b_i \rangle$ is zero in the Mott phase since
the particles are strongly localized and therefore removing a single
particle from a lattice site creates an orthogonal state. Conversely, in the
superfluid phase the system is in a coherent superposition of $n$-particle
states and therefore removing a single particle at one lattice site
does not significantly affect the overall state. This phase transition is experimentally accessible in an optical lattice setup \citep{Greiner2002} even in a two-dimensional system via the methods discussed previously.
Moreover, a frequency-dependent external perturbation that is well described using linear-response theory can be achieved via a small modulation of the lattice depth of $\lesssim 3\%$
close to the quantum phase transition \citep{Higgs:2012}. 
Thus, via a measurement of the temperature of the system in response to the external driving, the Higgs mode can be probed in the superfluid-Mott insulator transition.

\begin{figure}
\centering

\hspace{.05\textwidth}\includegraphics[width=.45\textwidth]{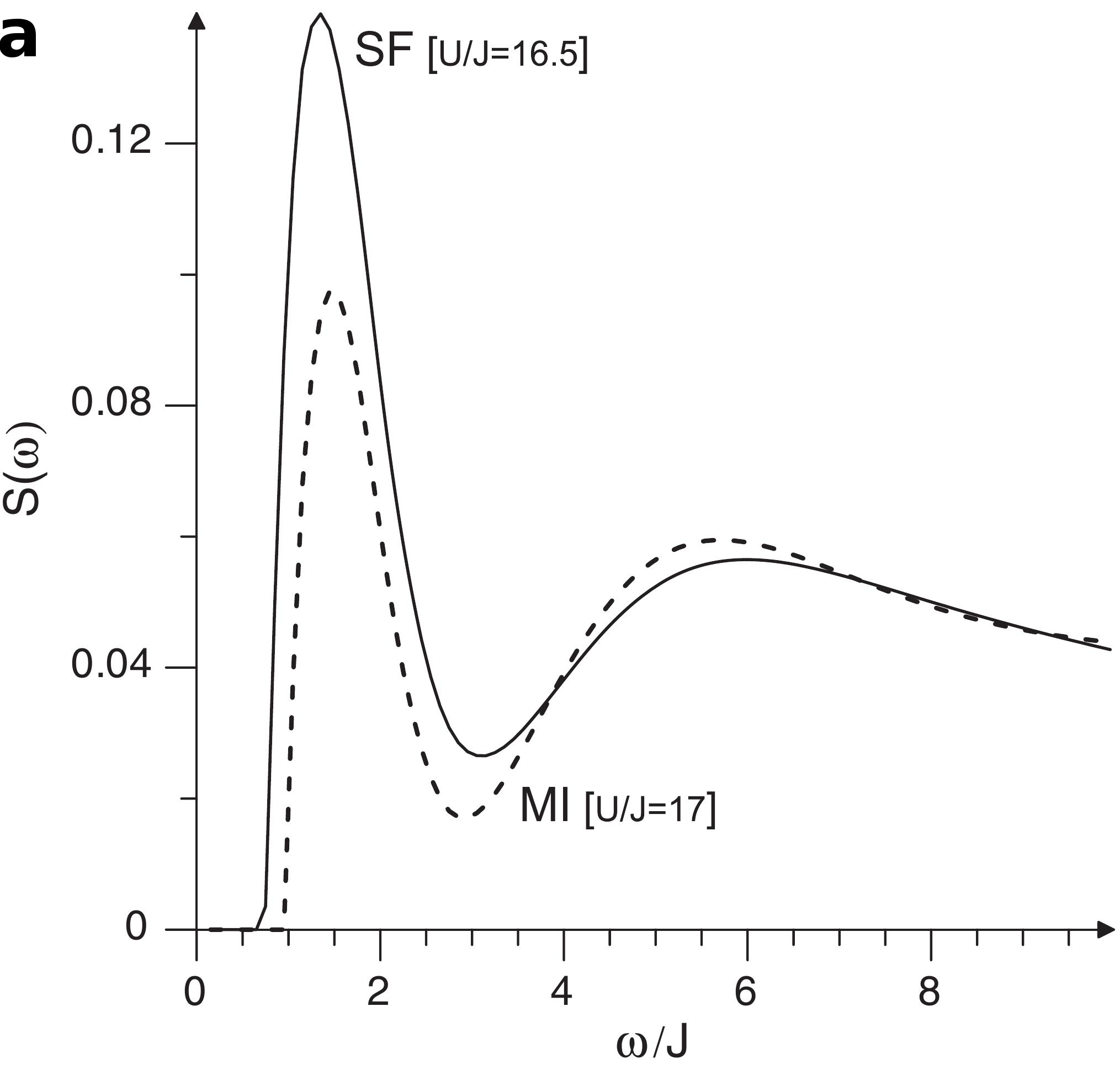}\hfill
\includegraphics[width=.35\textwidth]{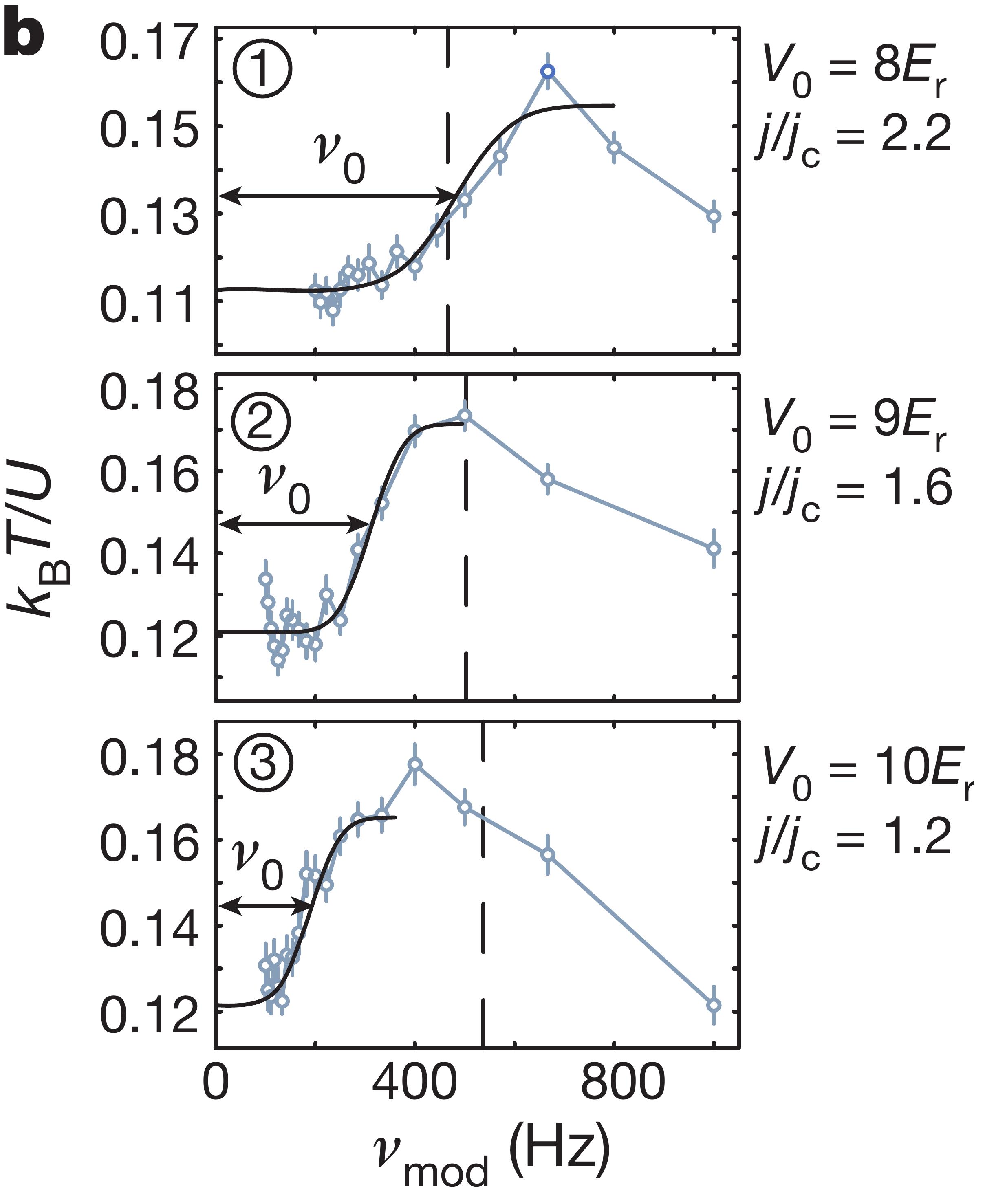}\hspace{.05\textwidth}
\caption{
(a) Expected resonant feature in the spectral response \citep[][Fig.~4]{pollet_higgs_2012}. 
(b) Experimental data for a fixed value of $j = J/U$ \citep[][Fig.~2]{Higgs:2012}. It is apparent that the resonant-like feature present in (a) cannot be observed in the experimental data (b).  
\label{fig:higgs_data} 
    }
\end{figure}

\paragraph{Experimental findings}

In the experiment reported by \citet{Higgs:2012} the authors measured the temperature response to external driving 
as a function of the lattice modulation frequency $\nu$. 
The experimental data exhibit a sharp spectral response at frequencies $\nu_0(j)$ that are in quantitative agreement with the analytical predictions for the gap of the Higgs excitations (see. Fig.~\ref{fig:higgs_data}(a)).
Moreover, as a function of $j$ they observe the characteristic softening of this gap
when approaching the critical point $j_c$ both from the superfluid and the
Mott insulating phase. However, the experimental data did not feature the expected resonance that is the `smoking-gun' signature of the Higgs mode. 
This might be due to effects of the harmonic confining potential \citep{pollet_higgs_2012,liu_massive_2015} and might be overcome by limiting the driving only to a small region in the center of the trap where the trapping potential can be approximated by a flat one \citep{liu_massive_2015}. 

In summary, while not fully conclusive yet, the data obtained in the quantum simulation of a relativistic $O(2)$-symmetric field theory is compatible with the existence of a collective Higgs amplitude in such a theory. 
Indeed, given the data, the authors were able to pick out certain theoretical predictions that correctly reproduce this data. 
This is a nontrivial task: many of the available predictions were conflicting since they were based on approximations, perturbation theory or numerical data in an attempt to tackle this computationally intractable problem. 
The experiment is compatible with and may be considered to suggest the existence of a gapped response and thus a Higgs mode given the (later) evidence that the broadening of the resonant peak can be explained by the inhomogenieties in the experimental setup. 
Moreover, this example of analogue quantum simulation unquestionably allows us to \textit{understand} features of $O(2)$-symmetric field theory by manipulating the ultracold atom source system.  

\subsection{Photonic Emulation and Quantum Biology}
\label{emulation cs}

\paragraph{Summary and Schema} In contrast to simulation, emulation involves the manipulation of a source system to gain knowledge pertaining to features of a concrete physical system. For example, an analogue quantum emulator may allow scientists to estimate salient spectral properties of complex molecules \citep{Huh:2014vk} or parameter regimes enabling phase transitions in exotic quantum materials \citep{Islam:2011ct}. 
In the past decade a vast body of experimental and theoretical work has emerged, demonstrating that quantum mechanical coherences (viz.\ the coherent addition of probability amplitudes) plays a critical role in many biological and chemical processes \citep{Scholes:2017gw}. Our second case study is based upon a formal analogy between single particle transport effects in photosynthetic complexes and photonic waveguide emulators. 

\begin{figure}
\begin{center}
\includegraphics[width=1.0\textwidth]{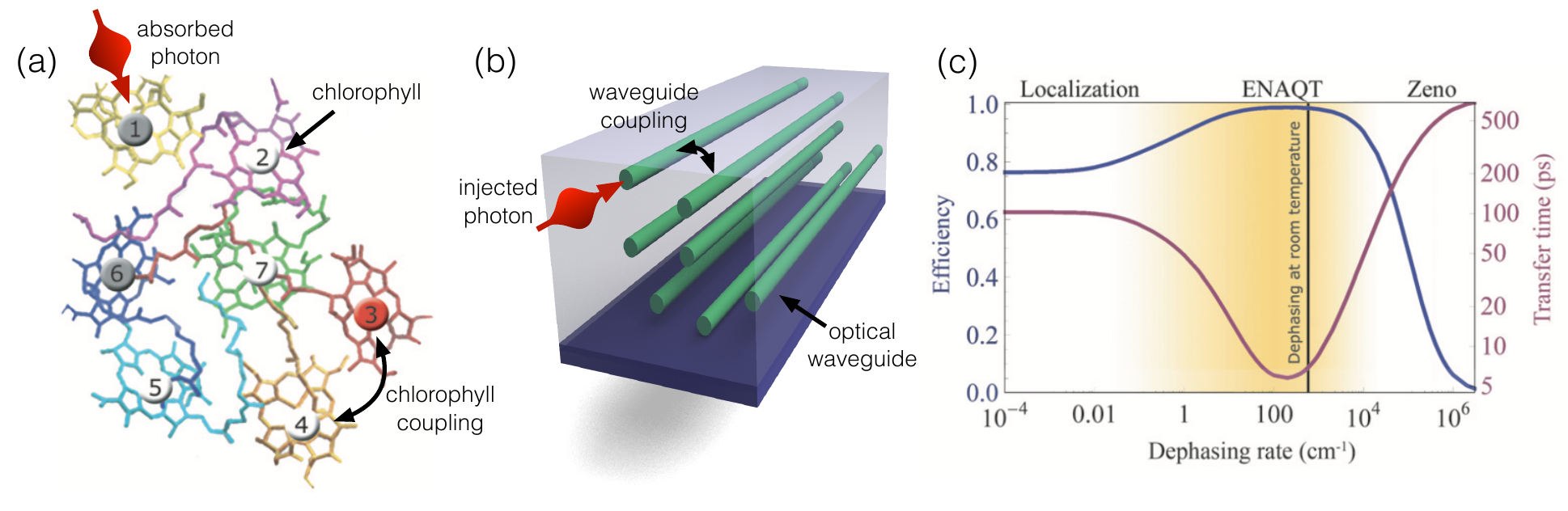}
\caption{Emulation of environment-enhanced quantum (ENAQT) transport in photosynthetic complexes.  (a) One of three FMO complex sub-units consisting of eight chlorophyl molecules (only seven shown) which enable exciton hopping between neighbouring sites. (b) A photonic waveguide emulator. (c) An optimal dephasing rate gives rise to a ENAQT.  Images (a,c) from \cite{Rebentrost:2009hu}.}
\label{fig:FMO}
\end{center}
\end{figure}

Biological processes appear to be not only robust too, but enhanced by environmental noise and disorder; running counter to the maxim that quantum coherences are easily destroyed. 
One particular phenomena that has received significant interest in recent years is environment-assisted quantum transport (ENAQT).  
ENAQT describes how the coherent transport of energy can be enhanced within certain regimes of environmental noise, and is proposed as an explanation for the exceptional efficiency of certain photosynthetic complexes.
For example, the Fenna-Matthews-Olson (FMO) complex is a protein complex found in green sulphur bacteria; an organism which can live in exceptionally low light environments. 
This complex is a trimer, with each sub unit consisting of eight chlorophyll molecules separated by a few nanometers [see Fig.~\ref{fig:FMO}(a)].
If a photon is absorbed by the light harvesting antenna (site 1), an exciton (i.e. an electron-hole pair) is transported via the neighbouring chlorophyll molecules towards the reaction centre (site 3) where a charge separation occurs and a biochemical reaction takes place. 
Whilst exceptionally long-lived coherences have been experimentally observed in photosynthetic complexes \citep{Engel:2007hb, Yang:2007kk}, understanding the role that coherences play in functionality is an outstanding challenge, and the subject of on-going experimental \citep{Panitchayangkoon:2010fw, Duan:2017cy} and theoretical \citep{Mohseni:2008gp, Rebentrost:2009hu, Wilkins:2015hv} research.

Photonic quantum technologies are systems which precisely generate, manipulate and detect individual photons \citep{Obrien:2009eu}.
Photons are appealing as a carrier of quantum information due to their inherent noise tolerance, light-speed propagation and ability to be manipulated by a mature integrated photonics platform \citep{Silverstone:2016gha}.
However, no quantum technology platform is without its drawbacks, and the inherent noise tolerance of photonics complicates the deterministic generation of entanglement which requires measurement and fast active feedforward \citep{Knill:2001vi}, or atom mediated interactions \citep{Lodahl:2015fy}.
Given this dichotomy, and owing to the ease of the high-fidelity manipulation of individual photon states, photonic quantum emulators are exceptionally well suited to exploring complex single particle dynamics, and have therefore recently emerged as a promising platform with which to explore ENAQT \citep{AspuruGuzik:2012ho}.
In the following sections we describe experiments which `programme' in an approximation of the Hamiltonian for the FMO complex into a photonic quantum emulator, and measure photonic transport efficiencies under certain models of artificially applied noise. 

The intensions of a scientist undertaking such photonic emulation seems two-fold: (1) to understand whether quantum coherences enhance functionality in real biological systems, and (2) whether understanding these effects can lead to technological breakthroughs in the development of new materials such as ultra-efficient photovoltaics \citep{Bredas:2017cq}.  The status of quantum functionality in biological systems is hotly debated\footnote{See \cite{Lambert:2012fj} for a balanced discussion.}, and in this context analogue quantum emulation has the potential to play a powerful inferential role. This case study illustrates our notion of analogue quantum emulation since features of the source system that is being manipulated (ENAQT into a photonic platform) are being appealed to for the specific purpose of gaining knowledge pertaining to features of actual and concrete physical system (ENAQT in a biological FMO complex). This is not a case of analogue quantum simulation since the intention of the experimenters is not to gain knowledge directly pertaining to features of an abstract theoretical model. 

\begin{figure}[t!]
\begin{center}
\includegraphics[height=0.3\textheight]{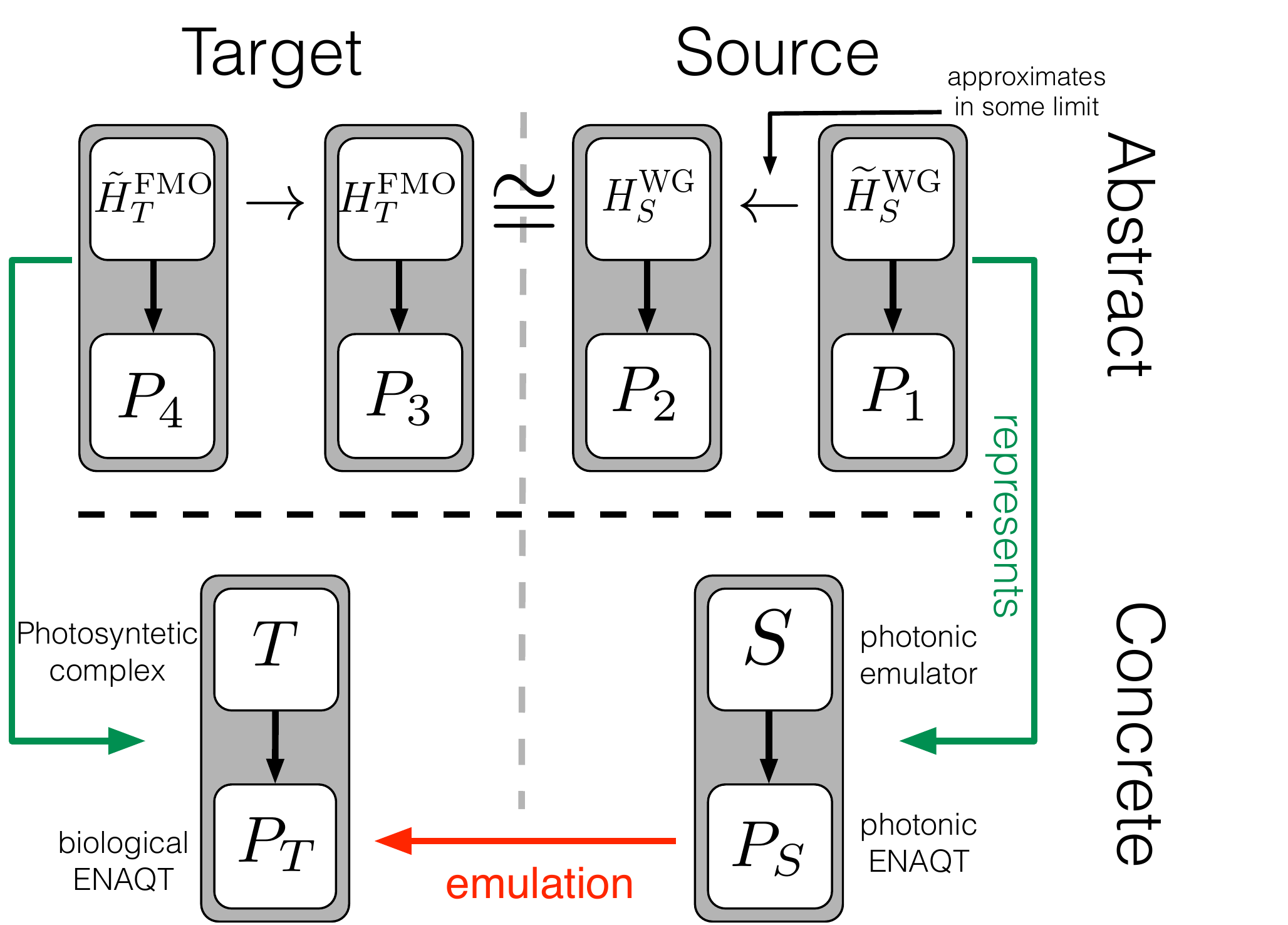}
\caption{Schema for Analogue Emulation Case Study (see main text for figure explanation).}
\label{fig:enaqt_schema}
\end{center}
\end{figure}

Let us use of our diagrammatic language to detail the various features of this case of analogue quantum emulation. The right hand side of  Fig.~\ref{fig:enaqt_schema} relates to the photonic waveguide source system. First we have some abstract Hamiltonian $\widetilde{H}^{\text{WG}}_S$ that corresponds to the concrete waveguide system, $S$ that includes experimental imperfections such as fabrication error, waveguide loss, dispersion and detector noise. This Hamiltonian can be approximated, in the appropriate limit, by the idealised waveguide Hamiltonian $H^{\text{WG}}_S$. 
The left hand side relates to the photosynthetic complex target system. The concrete target system, $T$, is the FMO complex and the phenomena $P_T$ is ENAQT.  In general, this system will be described by some complex Hamiltonian $\widetilde{H}_T^{\text{FMO}}$, which includes a non-Markovian phonon bath, relaxation effects and spatial correlations. 
It is then conjectured that the tight binding Hamiltonian $H_T^{\text{FMO}}$ approximates the target Hamiltonian $\widetilde{H}_T^{\text{FMO}}$ within some parameter regime such that the salient transport phenomena $P_T$ are sufficiently reproduced. Establishing the veracity of this conjecture is precisely the role of experimental and theoretical quantum chemistry.
$H_T^{\text{FMO}}$ in turn is isomorphic to $H^{\text{WG}}_S$. It is clear that $\widetilde{H}^{\text{WG}}_S$ stands in a representation relation with the photonic emulator source system, $S$, and $\widetilde{H}_T^{\text{FMO}}$ stands in a representation relation with the photosynthetic complex target system, $T$. Analogue quantum emulation is then a relationship between $S$ and $T$. The goal of analogue quantum emulation is to gain understating of actual phenomena in a concrete physical system. This is the key distinguishing feature between simulation and emulation that shall be the major focus of our analysis in the context of philosophical treatments of understanding in science.

\paragraph{Environment Assisted Quantum Transport}  Let us consider a specific example of biological ENAQT.  Following the analysis of \citet{Mohseni:2008gp} the FMO complex may be approximated by a tight binding Hamiltonian with $N=7$ sites:
\begin{equation}\label{eq:fmo}
	H^{\text{FMO}}_T = \sum_{m=1}^N \epsilon_m \ket{m} \bra{m} + \sum_{n<m}^N V_{m,n} (\ket{m}\bra{n}+ \ket{n}\bra{m}),
\end{equation}
where $\ket{m}$ represents an exciton at site $m$, $\epsilon_m$ the energy at site $m$ and $V_{m,n}$ the hopping potential between sites $m$ and $n$ (due to Coulomb interaction or electron exchange).
In this simplified model it is sufficient to consider a conserved single exciton tunnelling between sites as the recombination lifetime of the exciton (i.e. time until the exciton is lost) is significantly longer than the relaxation time of the chlorophyll (i.e. time taken to go from a high energy to low energy state). 

In multichromophoric arrays, coupling to a fluctuating protein and solvent environment via the electron-phonon interaction, induces time dependent variations in on-site energies and an irreversible dephasing of coherences; effectively describing the state by classical probabilities (real numbers) rather than quantum probability amplitudes (complex numbers).
Under certain assumptions\footnote{Specifically, that the phonon correlation times are short compared to the relaxation lifetimes, and that fluctuations at different sites are uncorrelated.} this site dependent dephasing can be described by a site independent pure dephasing rate, which can be used with the Lindblad master equation to describe the evolution of the system in the presence of environmental noise. 
Changes in dephasing rate are typically caused by variations in temperature.
In the study of environment-assisted quantum transport (ENAQT), scientists analyse the effect of the dephasing rate in transitions from a given input site [e.g.\ $m=1$ Fig~\ref{fig:FMO}(a)] to a given output site ($m=3$).
\cite{Rebentrost:2009hu} show that in the limit of zero-dephasing, i.e.\ purely coherent exciton hopping, variations in on-site energy restrict exciton transfer due to coherent interference between paths, a phenomenon known as Anderson localisation \citep{Anderson:1958fz}. 
This same effect causes sugar water to appear opaque, even though microscopically it is transparent to light.
As the temperature increases, dephasing disrupts this coherent interference; effectively unsticking the exciton and enabling transfer. 
However, if the temperature rises further, dephasing destroys all coherences within the system which, analogous to a quantum Zeno effect, suppresses transport.
\cite{Rebentrost:2009hu} show that at room temperature an optimal dephasing rate exists whereby transport efficiency is maximised towards unity, the so called `quantum Goldilocks effect' \citep{Lloyd:2011wg} [see Fig.~\ref{fig:FMO}(c)].

\paragraph{Photonic quantum emulators} 

Photonic quantum emulators typically comprise arrays of single mode waveguides, consisting a high refractive index core surrounded by a lower refractive index cladding. This set-up effectively confines the light and allows the construction of on-chip optical wires.\footnote{Precisely the same operating principle which enables fibre optical communication and long distance communication between Baleen Whales \citep{Payne:1971kr}.}
Connections between waveguides is achieved via evanescent coupling, whereby waveguides are bought close to one another such that the evanescent fields of the modes overlap, enabling photon tunnelling between neighbouring waveguides \citep{Politi:2008tl}.
Given a particular configuration of $N$ coupled waveguides, the system is described by the Hamiltonian
\begin{equation}\label{eq:wg}
	H^{\text{WG}}_S = \sum_{m=1}^N \beta_m \ket{m}\bra{m} + \sum_{n<m}^N C_{m,n}(\ket{m}\bra{n}+\ket{n}\bra{m}),
\end{equation}
where $\beta_m$ is the propagation constant for waveguide, $m$, determined by the refractive index of the mode; and $C_{m,n}$ is the coupling between waveguides, $m,n$, determined by the geometry and separation of the waveguides.
A single photon injected into mode $m$ evolves via $\ket{\psi(t)} = \exp(-i H^{\text{WG}}_S t)\ket{m}$, where time $t$ is related to the length $z$ of the coupling region via $z=ct/n$, where $c$ and $n$ is speed of light in a vacuum and the refractive index of the material respectively.
A further advantage is afforded by the fact the Schr\"{o}dinger equation is a wave equation, and therefore single particle dynamics can be simulated by injecting bright laser light.
Systems which exhibit coherent hopping between connected sites are known as quantum walks, and capture a very general class of phenomena \citep{Kempe:2003df}.

Critically, the isomorphism between equations \eqref{eq:fmo} and \eqref{eq:wg} means the task of building an ENAQT photonic emulator is two-fold: (T1) engineering the appropriate couplings between waveguide modes and (T2) engineering on-site dephasing. 
Recently, two complementary experiments were performed which addressed these tasks in turn.
\cite{Biggerstaff:2016bj} addressed (T1) by leveraging femtosecond-laser direct writing technology, which directly draws three-dimensional waveguides into glass, therefore enabling arbitrary couplings $C_{m,n}$ [see Fig.~\ref{fig:FMO}(b)].
The second approach by \cite{Harris:2017hi} lithographically patterns thermo-optic phase shifters on top of the waveguide circuit allowing them to control on-site dephasing (T2).
By engineering the magnitude of this dephasing, they observe both the peak and fall off in transport efficiency: the full quantum Goldilocks effect.

Let us make some remarks on these results. 
Each ENAQT quantum emulator has its respective advantages and drawbacks: the 3D waveguides of \cite{Biggerstaff:2016bj} enables arbitrary coupling between sites, but lacks the active control necessary for arbitrary dephasing; the lithographically fabricated waveguides of \cite{Harris:2017hi} enables high levels of control, but is inherently limited to one-dimensional connectivity.
Future ENAQT quantum emulators may use some combination of the two technologies to more closely mimic biological structures \citep{Smith:2009hm}, or, by engineering coupling to on-chip phonon modes \citep{Merklein:2017df}, emulate a realistic protein environment. 
In terms of scaling, each system emulates single particle dynamics and can therefore be modelled by classical wave dynamics (such as water waves).  
Therefore, these analogue quantum emulators provide no more than polynomial computational speedup over emulation on a classical machine.
Notwithstanding, multi-photon quantum walks have become an interesting and active research line \citep{Peruzzo:2010tq,carolan_experimental_2014, Carolan:2015vga}, and are closely related to the boson sampling problem which proves that mimicking the dynamics of a many-photon state is intractable on a classical machine (assuming a few reasonable conjectures).
Mapping this exponential speed-up onto a useful physical system is an outstanding open question, but recent theoretical evidence has suggested that a modification to the many-photon input state--- alongside Hamiltonians of the form \eqref{eq:wg}---enables the calculation of the vibronic spectra of molecules \citep{Huh:2014vk}, an important problem in quantum chemistry.



\section{Understanding Understanding}
\label{interpretive}

When we say that a particular theory, model or experiment provides a scientist with understanding of a physical phenomena, what do we mean? The standard philosophical strategy for addressing such questions is to attempt to formulate plausible necessary and sufficient conditions. Take the example of understanding via a model.  Most important accounts of understanding physical phenomena via models \citep{Friedman-JP-1974,Kitcher-PS-1981,trout_scientific_2002,Strevens2008,Strevens2013understanding,DieksDeRegt:2005} have the common feature that they include following three (necessary) conditions to be fulfilled \citep{ToyModels}: (i) the model must yield an explanation of the target phenomenon (explanation condition); (ii) that explanation must be true (veridicality condition); and (iii) the scientist who claims to understand the phenomenon via the model must have epistemic access to that explanation (epistemic accessibility condition). A useful account of understanding is Michael Strevens' simple view \citep{Strevens2008,Strevens2013understanding} according to which
\begin{quote} An individual scientist, $S$, understands  
    phenomenon, $P$, via model, $M$, iff model $M$ explains $P$ and $S$ grasps $M$.
    \citep[p.~17]{ToyModels}, 
\end{quote}

A central element of this notion of understanding is `grasping'. What is this supposed to mean? In Strevens' original account, grasping is (rather unsatisfactorily) posited as primitive: grasping is `a fundamental relation between mind and world, in virtue of which the mind has whatever familiarity it does with the way the world is' \citep[p.\ 511]{Strevens2013understanding}. Strevens thus adopts a primitive notion of epistemic accessibility in terms of a `subjective component' \citep[p.\ 122]{Bailer-Jones-1997} of understanding that cannot be reduced further. This is deeply unsatisfactory from our naturalistic view point and means that Strevens' grasping based account cannot be accepted as it stands. Rather, following \citet{Bailer-Jones-1997} and \citet{ToyModels} we will adopt a naturalistic account of this subjective component. That is, on our view what physical processes underlie 
grasping should be studied via cognitive science. 
For example, that a scientist grasps a model could mean that they construct a
corresponding `mental model' using which they can reason about the target
\citep{Bailer-Jones-1997, Bailer-Jones-2009}. 
In cognitive science mental models
are a model for cognitive processes, in particular, reasoning and knowledge
representation \citep{Nersessian:1999}. 
What is important for the philosophical discussion, however, is that there exists some
notion of grasping on which individual scientists can introspectively reflect. Think of the most basic and (arguably) fundamental physical model: simple harmonic motion (SHM). It seems to us unquestionable that any basic training in physics involves the process of `grasping' the model of SHM in the sense that the student acquires some form of mental model corresponding to a pendulum-like process. The acquisition of such a model is something we take to be both introspectively available to individuals and (in principle) externally analysable via cognitive science.   

On this account, if a scientist wants to obtain understanding of a phenomenon $P$ via an analogue quantum simulation (or emulation), the quantum simulation (or emulation) has to permit them to grasp the processes bringing about the phenomenon $P$. 
Plausibly, grasping requires at the very least that the dynamics of the simulator be observable in sufficient detail and manipulable to a sufficient degree. 
This is because only by being able to observe and manipulate the processes pertaining to $P$ can the scientist obtain the mental grasp of $P$ that is required for understanding. 
Plausibly, this is the major motivation for laboratory classes in undergraduate physics courses: students are able to observe key phenomena and manipulate the experimental systems in order to obtain a mental grasp and hence an understanding of the physics underlying those phenomena.
We therefore take it that observability and manipulability in this sense can plausibly function as an epistemic accessibility condition sufficient for establishing grasping.

As well as the notion of grasping the simple view of understanding makes use of the concept of explanation. This is a thorny and contested topic in the philosophy of science, with a long and largely inconclusive literature \citep{sep-scientific-explanation}. The simple view of understanding has the favourable property that it leaves open which accounts of explanation to favours in a theory of understanding. Moreover, the simple view has the added benefit of allowing us flexibility as to the role of truth (yet another thorny and contested topic!). In particular, we can make use of the distinction between `how-actually' and `how-possibly' explanations to define two different `modalities' of
understanding.\footnote{For more on how-actually and how-possibly explanations in general see \cite{dray:1968,Hemple:1965,reiner:1993,Forber:2010}. For a discussion specifically related to quantum computation see \cite{Cuffaro:2015}.} That is, whether or not the explanation via which $S$
understands is true allows us to distinguish between: (i) how-actually explanation that is required to be true; and (ii) how-possibly explanation is not required to be true. Following \cite{ToyModels},\footnote{For the sake of 
clarity, here, we neglect the contextual nature of understanding.} one can refine the above simple view of understanding accordingly:
\begin{enumerate}
\item A scientist S has \textit{how-actually understanding} of phenomenon $P$ via model $M$ iff model $M$ provides a how-actually explanation of $P$ and $S$ grasps $M$.
\item A scientist $S$ has \textit{how-possibly understanding} of phenomenon $P$ via model $M$ iff model $M$ provides a how-possibly explanation of $P$ and $S$ grasps $M$.
\end{enumerate}  

Let us give two brief examples to illustrate the distinction between how-possibly and how-actually understanding. Schelling's model of residential segregation yields merely how-possibly understanding of the actual segregation in Chicago in 1968. The model assumptions---that Chicago is a checkerboard, that residents can be reduced to their skin colour, etc.\---are too unrealistic to plausibly yield any understanding of the concrete and actual target phenomenon. In contrast, the Heisenberg model in physics might yield how-actually understanding of magnetism based upon quantum physics and the realistic assumption that the magnetic dipole moment can be modelled as a three-component spin. Hence, the kind of understanding we can obtain from the Heisenberg model is how-actually understanding about how magnetism arises in actual solid states.  In practice, the  how-possibly vs.\ how-actually explanation distinction, and thus the how-possibly vs.\ how-actually understanding distinction, might plausibly be taken to come in degrees. 
That is, since one might speak of the degree to which a model is true,\footnote{For example see \cite{smith:1998}.} one might consider a modally variating spectrum of understanding ranging between the two extreme cases. 
Such subtleties can reasonably be neglected for the purpose of this analysis. 
Rather, it will prove instructive to not only retain a binary distinction between how-possibly and how-actually understanding, but supplement it with further binary distinction, that between concrete and abstract phenomena.

Consider again our schema for understanding.  
Clearly the phenomenon $P$ in question need not be physically instantiated in any concrete physical system. Instead we could take the phenomenon about which the scientist is gaining understanding via a model to be `abstract target phenomena'. 
By how-actually understanding of an abstract target phenomenon we have in mind examples such as `frustration' in magnetic spin networks.  Here, networks of spins are connected via anti-ferromagnetic interactions, which attempt to orientate neighbouring spins towards opposite directions.   Under certain lattice geometries (for example, a three particle triangular geometry), no spin configuration can satisfy all interactions.
How actually understanding of this abstract target phenomena requires exploring this target in different parameter regimes via experimental manipulation (i.e.\ simulation \citep{Kim:2010ib}) or numerical modelling.
In some cases, of course, the phenomenon will be physically instantiated and thus the relevant understanding is of `concrete target phenomena'. 
For example, the abstract phenomenon of frustration has been shown to be central in the understanding of protein folding \citep{Bryngelson:1987uu}.

Just as one may have how-possibly or how-actually understanding of concrete phenomena, one may also have how-possibly or how-actually understanding of abstract phenomena.\footnote{The veridicality condition for understanding of abstract phenomena via models might be  satisfied, if the explanatory model is an \emph{embedded} model in the sense of \citet{ToyModels}. Indeed, understanding via toy models seems to be a particular instance of understanding abstract target phenomena as we detail below (Sec.~\ref{map}. } Thus, we are left with four types of understanding distinguished by our two binary distinctions. In the following section we will uses these four cases to define a `methodological map' that situates analogue quantum simulation and analogue quantum emulation alongside traditional forms of scientific activity. In Section \ref{normative} we will then link the simple account of understanding to the actual scientific practice in that we discuss concrete norms that must be met for understanding via analogue quantum simulation and emulation to be achieved. 

\section{Methodological Mapping}
\label{map}

\begin{figure}[t]
\centering
\includegraphics[width = \textwidth]{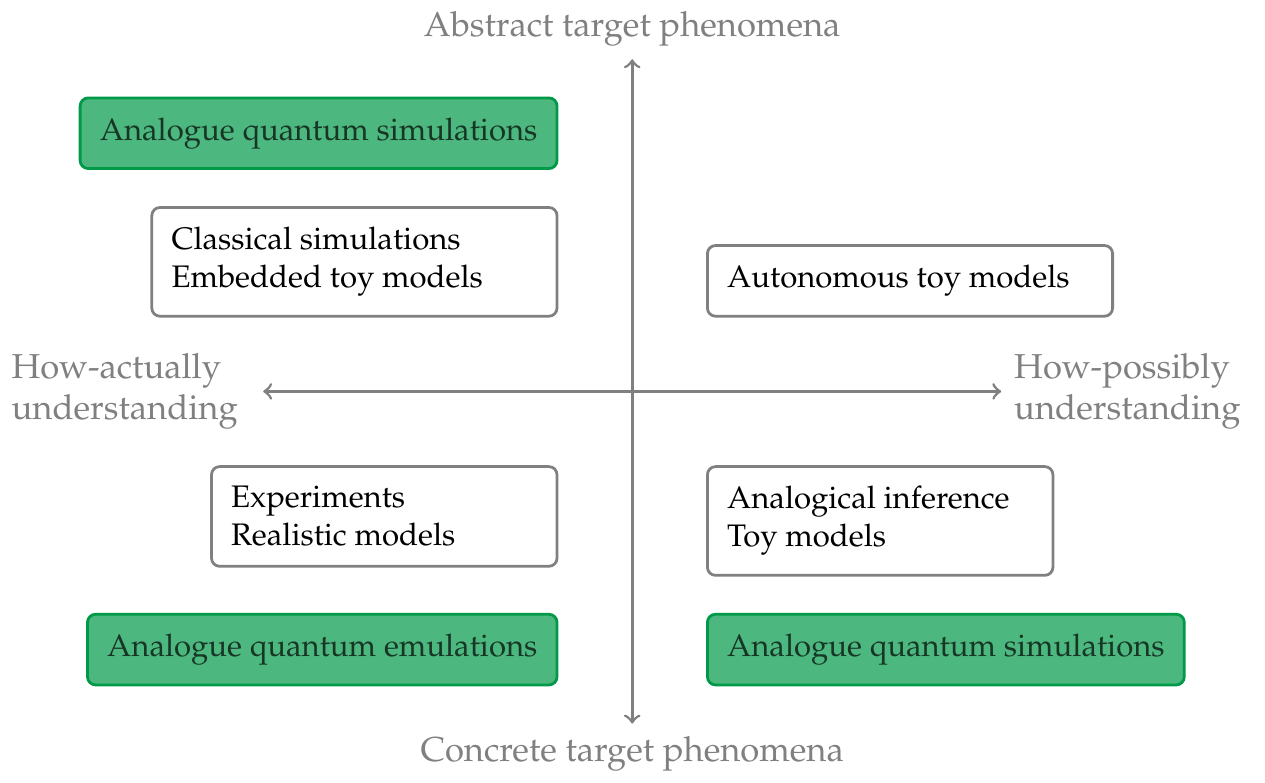}
\caption{Methodological map with respect to the type of understanding aimed at and the kinds of phenomena that are to be understood.} 
\label{fig:methodological map}
\end{figure}

The guiding principle of this part of our analysis is that the goals a scientists has in using a particular form of scientific inference are usefully characterised by the form of understanding that they expect to acquire. We will first illustrate this idea with reference to traditional forms of scientific inference, before returning to analogue quantum simulation and analogue quantum emulation. The results of our analysis are summarised in Figure~\ref{fig:methodological map}. It is not our assumption here that understanding is the only goal of science but rather that consideration of this goal is sufficient to uncover the principal methodological differences at hand.\footnote{For a recent debate on the epistemic value of understanding and knowledge see, for example, \citet{kvanvig_value_2003} and \citet{pritchard_knowledge_2014}. For discussion specifically focusing on understanding as the goal of science see \cite{rowbottom:2015,bangu:2015,dellsen:2016,stuart:2016,park:2017}.}  

Inferences built upon experiment are the gold standard of scientific reasoning. If the account of scientific understanding considered in the previous section is to have any general applicability to science then surely it must allow us to characterise the form of understanding that scientists hope to gain in carrying out an experiment. Let us consider arguably the most famous \emph{experimentum crucis} 
of the twentieth century: Eddington's 1919 measurement of the deflection of optical starlight as it passed the sun during a solar eclipse \citep{kennefick:2009,will:2014}. A simplified reconstruction of the reasoning that this experiment entailed runs as follows. Lets us designate the Schwarzschild solution to the Einstein field equations the model, $M$. The concrete phenomenon, $P$, is the deflection of light by a gravitational field. $P$ is a logical consequence of $M$ together with the appropriate auxiliary assumptions. Furthermore, irrespective of ones account of explanation, it is clear that $M$ should be counted as a putative explanation for $P$.  The Eddington experiment provides an experimental exemplar of $P$ as mediated by $M$. Given this, $M$ provides a how-actually explanation of $P$. That is, the Schwarzschild solution provides a how-actually explanation for the deflection of light by a gravitational field. In employing $M$ in such an explanation,  (plausibly) a scientist, $S$ would `grasp' the relevant aspects of $M$. We can assume that Eddington had some cognitively accessible mental model corresponding to the relevant aspects of the Schwarzschild solution. Putting everything together: Eddington's goal in carrying out the 1919 experiment was to gain how-actually understanding of the concrete phenomenon of the bending of light by a gravitational field via appeal to the Schwarzschild solution. He grasped the relevant aspects of $M$ and wanted to ascertain whether $M$ provided a true explanation of $P$. In general terms, we can characterise the understanding that scientists hope to gain in carrying out experiments as being how-actually understanding of concrete phenomena.\footnote{This is not to say that this is the only thing that scientists hope to gain in carrying out experiments or that all experiments will lead to how-actually understanding.}  

Next, let us consider the form of understanding that scientists aim for in employing simulation via a classical digital computer. The explanatory reach of computer simulations with regard to material systems is a matter of controversy in the literature. On the Parker-Winsberg account (WPA), we take computer simulations to be able to provide us with insight into concrete phenomena, much like an experiment \citep{parker:2009,winsberg:2009,Winsberg-2010,Winsberg-SEP-2013}. In this case the goal of scientists in carrying out computer simulations would be to provide how-actually explanations of concrete phenomena and thus to (potentially) provide how-actually understanding of such phenomena. Such an account would seem to fit well with, for example, the explanation of the phenomenon of galactic tails and bridges provided by the computational model of their formation via tidal forces in galactic collisions \citep{Toomre:1972}. Contrastingly, on the Beisbart-Norton account (BNA) \citep{beisbart:2012}, computer simulations are simply `arguments' and can only tell us about solutions to equations in a manner inferentially identical to pen and paper calculation. 
In this case the goal of scientists in carrying out computer simulations would be to provide how-actually explanations of abstract phenomena. 
Provided the grasping condition is met, one would then take computer simulations to provide how-actually understanding of abstract phenomena. 
The Beisbart-Norton account seems to fit very well with the use of computer simulations in the numerical integration of equations of motion or in Monte Carlo simulation. For example, in employing a computer simulation to solve an $N$-body problem in Newtonian particle mechanics, one plausibly gains how-actually understanding of stable and generic features of the solutions, such as Kepler pairs. 

Analogical inferences have long played an important role in science and the topic has been fairly extensively discussed in a literature that features work
by \citet{keynes:1921},  \citet{hesse:1964,hesse:1966}, \citet{Bailer-Jones-2009} and \citet{bartha:2010,bartha:2013}. 
An
important distinction, due to \citet{hesse:1964}, is between `material
analogies', that are based upon to relevant similarity of properties between two systems, and
`formal analogies', that obtain when two systems are both `interpretations of the same
formal calculus' \citep{Frigg:2012}. 
In both cases, the analogical relationship
holds between the source and target systems in question, rather than a model understood to represent these systems in some restricted domain of validity. 
Following Hesse, we should thus take both formal and material modes of analogical inference to be forms of reasoning that relate to concrete rather than abstract phenomena. 
Analogical
reasoning of the type discussed by Hesse typically takes the form of a speculative inference. 
Classic examples are Reid's argument for the
existence of life on other planets \citep{reid:1850} and Hume's argument for
animal consciousness \citep{hume:1738}. 
Authors such as \cite{salmon:1990} and \cite{bartha:2010,bartha:2013} take arguments by analogy to establish only the \textit{plausibility} of a conclusion, that is, that there is some reason to believe in that conclusion, and with it grounds for further investigation. 
Given this, the form of explanation that analogical inferences can be expected to provide is restricted to the modally weaker how-possibly form. 
Provided that the scientists grasps the formal or material analogy in question, it is thus plausible to take arguments by analogy to provide how-possibly understanding of concrete target phenomena.  

Toy models are highly idealised and simple tractable models. 
They can be contrasted with realistic models that involve a large number of modelling assumptions and parameters used to most accurately describe some concrete target system. 
Compare, for instance, the Ising model in physics with climate-models. One core aim of both forms of modelling is to obtain some kind of understanding of a target phenomenon \citep{ToyModels,Hangleiter-2014,DieksDeRegt:2005}.\footnote{That is not to say this is the only aim. See \cite{Sugden-JEM-2000,Niss-AHES-2011,Hangleiter-2014,Frigg:2012}.} \cite{ToyModels} argue that toy models are studied precisely because they: (i) permit an explanation of abstract target phenomena, that might in some cases be relevantly related to a true explanation; and (ii) are simple enough so that individual scientists can grasp them. 
In particular, the simplicity of toy models facilitates analytical solutions or simple computer simulations using which we can gain a grasp of their abstract target phenomena. 
This may take the form of an intuition about `What if?' questions, or of appropriate mental models.
Following the argument of \cite{ToyModels}, toy models can be categorised into embedded toy models, that is, `models of an empirically well-confirmed framework theory' (p.\ 4), and autonomous toy models that are not embedded. 
On this account autonomous toy models are employed with the aim of providing how-possibly understanding of abstract target phenomena.
In contrast, embedded toy models are employed with aim of providing how-actually understanding of abstract target phenomena. 

Our main purpose in pursuing the foregoing analysis is to enable us to situate analogue quantum simulation and emulation on the `methodological map' of modern science. The emulation case is the most straightforward. Recall our case study: There, the goal of the scientists is clearly to provide an exemplar that supports the claim that ENAQT provides a true explanation for concrete phenomenon in biological systems. Emulation is, in this regard, much like experiment: it is an inferential tool aimed at probing a class of concrete phenomena by manipulating a concrete system. Given that the scientists carrying out an emulation grasps the models that are being employed (${H}^{\text{FMO}}_S$ and ${H}^{\text{WG}}_S$ in our example), we can say that the analogue quantum emulators are employed to provide how-actually  understanding of concrete target phenomena.  The case of analogue quantum simulation is more ambiguous. Evidently the function of such simulations has much in common with that of computer simulations. Thus it seems very plausible to say that analogue quantum simulators, like the cold atom simulator considered in our example, are employed to provide how-actually understanding of abstract target phenomena, such as the 2D Higgs mode in $O(2)$-symmetric field theory. 
It is also clear, however, that there is a strong parallel between the function of analogue quantum simulators and that of analogical inference. 
Both appear well suited to provide modally weaker forms of understanding of concrete phenomena such as establishing the plausibility of a conclusion.\footnote{Immanuel Bloch and Ulrich Schneider, both scientists working with such simulators, in fact expressed precisely this sentiment in field interviews conducted towards this project.} 
We thus further claim that analogue quantum simulators are employed to provide how-possibly understanding of concrete target phenomena. 
Consequently, we can situate analogue quantum simulation and emulation on the methodological map as shown in Fig.~\ref{fig:methodological map}.

\section{Norms for Simulation and Emulation}
\label{normative}

\subsection{Epistemic Norms}

Above we analysed the goal of experimental science in terms of the achievement of how-actually understanding. The key idea was that an experiment provides an exemplar of a concrete phenomenon, $P$, sufficient to show that a particular model, $M$, provides a true explanation of $P$. In order to assess the conditions for such a goal to be achieved it is evidently necessary to dig a little deeper into the epistemological foundations of experimental science.\footnote{See \citep{Franklin:2015} for a full review.} Following \cite{franklin:1989}, one of the key ideas in the epistemology of experiment is that to assess the inferential power of experimentation, we must examine and evaluate the strategies that scientists use to \textit{validate} observations within good experimental procedures. Following \cite{Winsberg-2010} we can draw the distinction between two different types of validation in the context of experimental science: an experimental result is `internally valid' when the experimenter is genuinely learning about the exemplar system they are manipulating; an experimental result is `externally valid' when the information learned about the exemplar system is relevantly probative about the class of systems that are of interest to the experimenters. So we see that the preconditions for us to gain how-actually understanding via an experiment clearly must include internal and external validation. This is directly analogous to the explanation condition (the explanation must be logically coherent etc.) and the veridicality condition (the explanation must be true) which must be met for understanding via a model according to the simple view (cf.\ Section~\ref{interpretive}). 

\begin{figure}
\centering
\includegraphics[height=0.35\textheight]{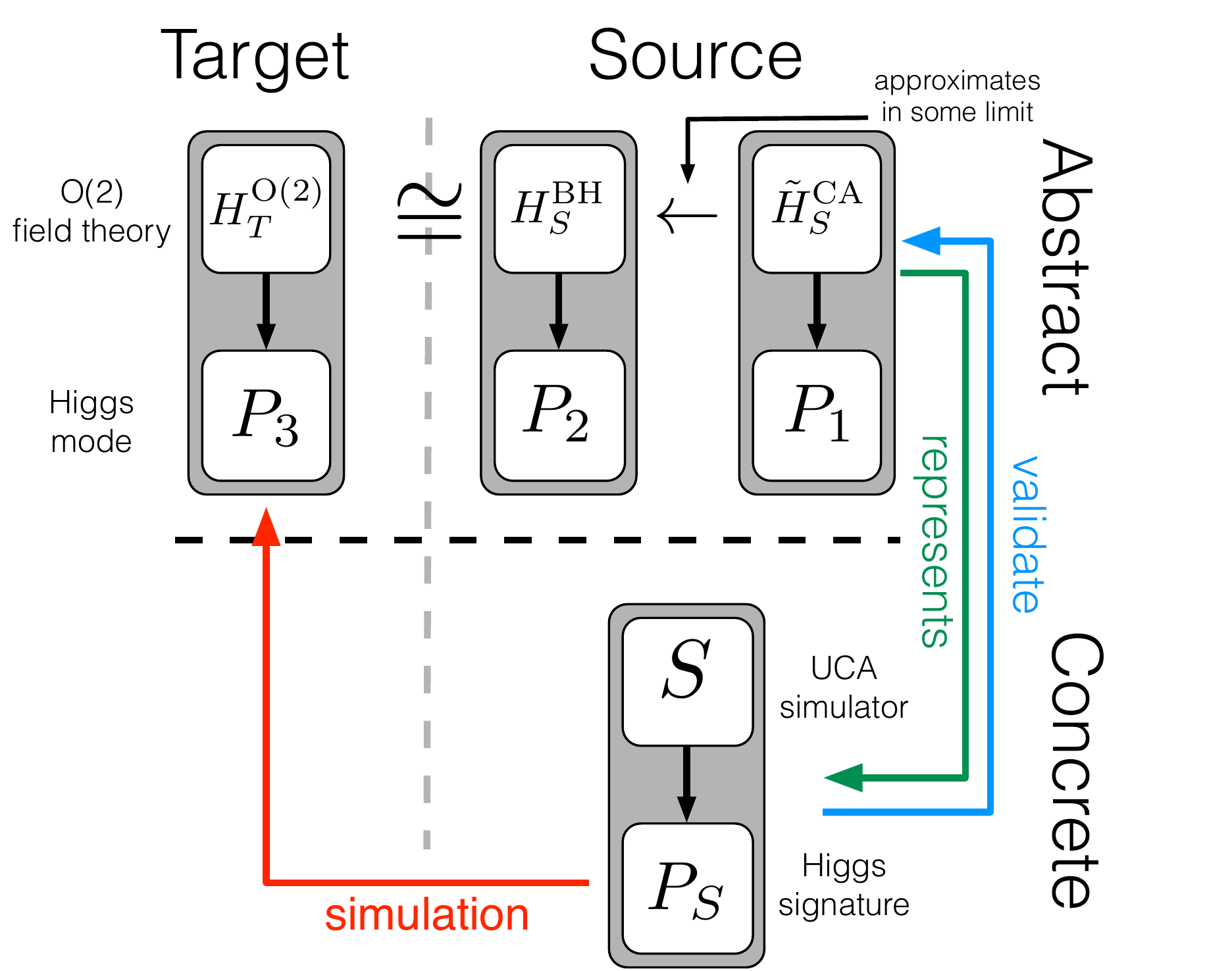}
\caption{
We show the schema for simulation of the Higgs mode in two dimensions via cold atoms in optical lattices, illustrating the steps that need to be taken for validation of the simulation. 
\label{fig:higgs scheme validate}}
\end{figure}

\begin{figure}
\centering
\includegraphics[height=0.35\textheight]{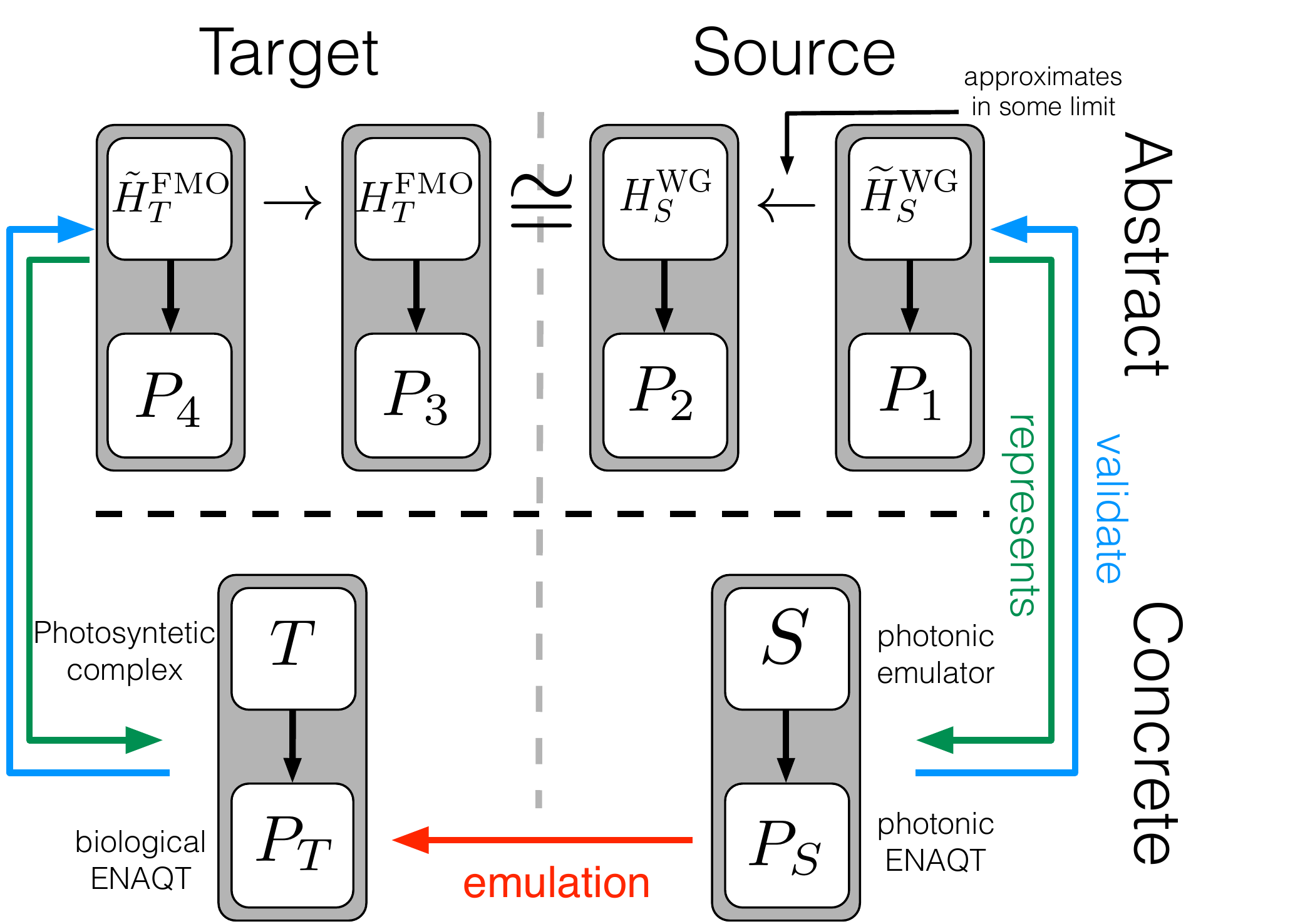}
\caption{
We show the schema for emulation of environment-assisted quantum transport via a photonic quantum simulator, illustrating the steps that need to be taken for validation of the emulation. 
\label{fig:enaqt scheme validate}}
\end{figure}

Such norms for the epistemology of experiment and understanding via models are naturally extended into the epistemology of analogue quantum simulation and emulation. 
The primary function of analogue quantum simulations is to compute features of
the target theory or model and thus obtain how-actually understanding of an abstract target phenomenon. 
This goal matches that of a computer simulation the goal of which, at least on Beisbart-Norton account, is to learn about features of one's theory that are not accessible by analytical means. 
The primary function of analogue quantum emulation is to obtain how-actually understanding of concrete target phenomena. In order for both analogue quantum simulators and emulators to be performing their proper function we need to certify the approximations required to connect the abstract target phenomenon with the source model in the first place (internal validation). 
This certification may proceed in two steps: first we certify the relationship between source model and target model and then we certify the relationship between target model and abstract target phenomenon (as illustrated in Figs.~\ref{fig:higgs scheme validate} and \ref{fig:enaqt scheme validate}). Moreover, for the case of quantum simulations, we need to validate the correspondence between abstract and concrete source systems. 
Otherwise we can never be sure that the simulator is in fact solving the relevant source model (internal validation). 
For quantum emulations we need to furthermore validate the correspondence between abstract and concrete target systems (external validation).  
Such manifold validation required for both simulation and emulation will likely be a collaborative process, requiring expertise spanning multiple scientific disciplines. 

Scientists are thus not successfully achieving the goal of quantum simulation or emulation unless all correspondence relations (source system--source model, source model--target model, and target model--target system as represented by blue arrows in Figs.~\ref{fig:higgs scheme validate} and \ref{fig:enaqt scheme validate}) can be convincingly established. An interesting consequence of the combinations of our definitions with our epistemic norms is that a given experiment may fail as an emulation and yet succeed as a simulation. That is, a scientist may carry out an experiment on a source system with the intention of gaining knowledge of regarding an actual and concrete physical system, achieve the internal validation necessary for simulation, and yet fail in their goal of emulation due to lack of external validation. Such considerations suggest that, as mentioned above, in practice real cases may need to be conceived of as having components of both simulation and emulation simultaneously. The crucial point is that validation procedures track the simulation vs.\ emulation distinction. So, although a case of emulation could always be plausibly reinterpreted as a simulation, the opposite would require an additional experimental step. 
More precisely, to transform a successful simulation into a successful emulation one needs external validation and this will invariably involve changes in the experimental protocols needed to validate the target model--target system correspondence. 

In general, certification of quantum devices, and hence internal validation of quantum simulators or emulators, is a burgeoning field \citep[for a recent review see][]{gheorghiu_verification_2017}. There, the fundamental difficulty lies in the speed-up that quantum devices offer over classical computers so that predictions are not easily checkable using numerical simulations on classical computers. 
Broadly, internal validation requires confirmation that phenomena $P_S\cong P_2$. 
In a quantum device, this might be done directly, invoking certain assumptions about ones experimental setup \citep[e.g.][]{Certification,takeuchi_verification_2017}, using some quantum capacities of the certifier \citep[e.g.][]{fitzsimons_post_2015,fitzsimons_unconditionally_2017-1,wiebe_hamiltonian_2014}, or via a process of building trust in the experimental setup by certifying it in computationally tractable regimes, reaching towards the edge of computational tractability \citep[e.g.][]{carolan_experimental_2014,BlochEisertRelaxation}.  
The idea behind the latter approach is to give the experimentalist confidence that the phenomena will pertain in the intractable regime where classical computational techniques can not longer be employed.

We can now illustrate these steps on the basis of the case studies on simulation and emulation. We start with the case of simulating the Higgs mode in two dimensions using cold atoms in optical lattices described in Fig.~\ref{fig:higgs scheme validate}. 
In order to (internally) validate $S$ as a simulator of $P_3$, the 2D Higgs mode, one needs to validate the representation relation between $S$ and $\tilde{H}_S^{\text{CA}}$ in the relevant parameter regime.
Simulation of the Higgs mode starts from a theoretically observed correspondence between $H_T^{O(2)}$ and $H_S^{\text{BH}}$ such that the precise parameter regime is known in which the Bose-Hubbard Hamiltonian realises an $O(2)$ field theory that exhibits or not a Higgs mode \citep{altman_oscillating_2002}. 
Here, this is the case in the vicinity of the SF--MI phase transition, i.e.\ at unit filling and when $j \sim j_c$, in which the experiment is performed. 
In a first step, one now needs to validate the fact that $S$ is well-represented by the Hamiltonian $\tilde{H}_S^{\text{CA}}$, a realistic model of the simulator system. 
In this model additional terms to those in the Bose-Hubbard Hamiltonian \eqref{bose hubbard} are present, for example, next-nearest-neighbour hopping terms and coupling to the environment. 
For this first step of validation, one needs to obtain precise values (estimates) for the strength of the individual terms of interaction present in $\tilde{H}_S^{\text{CA}}$. 
In a second step, given $\tilde{H}_S^{\text{CA}}$ as a good model of $S$ one can then rigorously show to what accuracy $\tilde{H}_S^{\text{CA}}$ approximates the Bose-Hubbard Hamiltonian $\tilde{H}_S^{\text{BH}}$ \citep{Jaksch1998}. 
This can be done by putting quantitive bounds on the signficance of the additional terms, given the estimates of the relevant interaction parameters \citep[see e.g.][]{BlochEisertRelaxation}.
The experiment of \citet{Higgs:2012} in fact only partially achieved this final goal since the experimenters failed to fully validate the Bose-Hubbard model as a good approximation to $\tilde{H}_S^{\text{CA}}$. 
Specifically, it seems to be the case that the harmonic confining potential that is not taken into account in the Bose-Hubbard Hamiltonian might have an impact on the signature of the Higgs mode in the cold-atom system (cf.\ Sec.~\ref{higgs cs}).

Let us next consider analogue quantum emulation. Here, in order for scientists to achieve their goal of how-actually understanding of concrete phenomenon we require not only all the forms of validation and certification relevant to analogue quantum simulation, but also something more: they must validate the correspondence between abstract and concrete source systems (internal validation) \emph{and} validate the correspondence between abstract and concrete target systems (external validation).  
In our case study internal validation takes much the same form as for the ultracold atom simulator.  
The photonic emulator $S$ is fully characterised using both classical and quantum techniques to determine relevant parameters such as waveguide loss, dispersion, light source parameters (brightness, photon indistinguishability) and detector parameters (dark counts, efficiency, shot noise).
This information is fed into a computational model of the emulator $\tilde{H}^\text{WG}_S$, and provided the errors lie within some bound (a notion which can be made theoretically rigorous) we establish the approximation relation $\tilde{H}^\text{WG}_S\rightarrow  {H}^\text{WG}_S$.
Validation of the representational relation between the abstract and concrete emulator is somewhat complicated by the fact the concrete system is inherently quantum, and full characterisation involves quantum tomography which requires a number of measurements that scales exponentially in the size of the system \citep{Nielsen:2011vx}.

External validation of the model of light harvesting complexes is an outstanding challenge in experimental and theoretical quantum chemistry.  Experimentally, 2D electronic spectroscopy is used to determine the structure of the molecule (e.g.\ the energy levels), and to observe long lived quantum coherences.  This in itself is only half the picture, as the context in which these experiments are performed (in vivo, in vitro, high/low temperature) is critical.
To validate the mode, computations are performed on $\tilde{H}^\text{FMO}_T$ to ascertain whether it reproduces known target phenomena (i.e. $P_4\cong P_T$). This in general will require approximations, and further work is required to establish the applicability of those.
If the approximation between $\tilde{H}^\text{FMO}_T\rightarrow {H}^\text{FMO}_T$ can be established (analytically or numerically) then the emulator may itself help validate the representation relation.  Finally, once the relationship is validated the emulation can be performed for unknown target phenomena.

The subjective component of understanding involved in analogue simulation and emulation is the hardest to analyse. 
Plausibly, as argued above, a minimal requirement that needs to be fulfilled to satisfy the epistemic accessibility condition is that the dynamics of the system be observable in sufficient detail and manipulable to a sufficient degree.
Indeed, only to the degree that $P$ and, equally important, the process bringing about $P$ is observable and manipulable does a simulation or emulation permit epistemic access to the explanation of $P$ and hence understanding of $P$ according to the simple view. 

Take the example of the quantum simulation of the two-dimensional Higgs mode. Here, the relevant measured quantity is the spectral response of the system to external driving. 
This is the output of the simulation in that the `smoking-gun' features of the Higgs mode are observable in terms of an onset of a resonant feature in the spectral response. 
Hence, a minimal requirement for grasping is satisfied. 
Nevertheless, to limitations in the experimental setup the onset of the spectral response could not be observed for reasons discussed by \citet{liu_massive_2015}.
What is more, neither could any other quantities be observed that might allow for a more detailed error analysis, nor was the process bringing about the purported resonant feature and thus the Higgs mode accessible. 
Judging by our epistemic norms, the experiment by \citet{Higgs:2012} seems by itself insufficient to establish understanding of the Higgs mode in 2D.
This notwithstanding, the analysis by \citet{liu_massive_2015} lends some \emph{a posteriori} justification to the experimentally observed lack of a resonant-like feature and suggests a path toward a more accurate simulation of the Higgs mode in 2D using optical lattices.
Summarizing, while not fully satisfactory, the experiment of \citet{Higgs:2012} and follow-up work has increased our understanding of the mechanisms dominating the SF--MI transition.

\subsection{Pragmatic Norms}

The norms of validation and certification for analogue simulation or emulation are crucial for its \emph{epistemic status} regarding the enterprise of obtaining understanding of a simulated or emulated phenomenon. It also seems reasonable to require more \emph{pragmatic norms} of a quantum simulation or emulation. Such norms  pertain to the \textit{usefulness} of a simulation towards goals that are not directly epistemic. Based upon our cases studies we suggest the following:   

\paragraph{Heuristic Norm: Observability}

One central use of analogue quantum simulations and emulations particular is as a heuristic for future theory construction.\footnote{This statement is also supported by our field interviews.} Such heuristic value of analogue quantum simulation seems naturally connected to the epistemic function as a means towards (how-possibly) understanding. 
In particular, the subjective competent of understanding, that we characterised via grasping, could reasonably taken to have a heuristic as well as an epistemic function.
In enabling scientists to grasp an abstract target model, an analogue simulation will provide a `mental model' that can be manipulated, iterated and refined. Such mental models may, in the sense of \cite{saunders:1993}, be \emph{heuristically plastic}.  As indicated above, a reasonable minimal condition for any such endeavour must be that the source system is observable and manipulable to a sufficient degree. 
One must be able to probe the simulator very precisely so that one is able to create an heuristically plastic model of the relevant processes. 
The heuristic value of an analogue simulation is related but distinct from its epistemic value in terms of how-possibly understanding of concrete target phenomena, it thus licences a separate pragmatic norm.

\paragraph{Computational Norm: Speed-up} 

A second pragmatic norm for quantum simulators and emulators is that they outperform the best possible classical device.
There are two notions by which we can mean `outperform'.
The first, and more traditional notion refers to an asymptotic scaling in the computational resources (e.g. time or space) required to compute a given function.
Whilst classical computers can calculate certain functions efficiently (i.e.\ a scaling in resources that grows at most polynomially in the size of the problem), many problems do not belong to this class of functions, in particular, the simulation of certain quantum mechanical phenomena \cite[e.g.][]{Kempe-SIAM-2006, Osborne-2011}. 
Building on the ideas of \citet{Feynman:1982,feynman:1986} it was proven by \citet{lloyd_universal_1996} that quantum systems can be efficiently simulated using other quantum systems, and thus `outperform' a classical device.
For example, in the case of the simulation of the Higgs mode in two dimensions (Sec.~\ref{higgs cs}), a classical computer is likely to require resources that grow exponentially in the size of the system. 

This notion of scaling is valid in the error free, or error corrected regime, where the effect of experimental imperfections will not overwhelm the accuracy of the computation as it is scaled-up.
However, experiments are not ideal, and to-date there is no known way to error correct analogue quantum devices \citep{Hauke:2012dq}.
This therefore brings with it a second, more restricted notion of `outperform', whereby for a given problem size (or limited range of sizes) our analogue quantum simulator has a runtime quicker than our best known classical algorithm, enabling the calculation of useful phenomena not otherwise possible.
This more pragmatic notion of `outperform' may also offer a deeper or more detailed epistemic access to the target phenomenon.

We consider the following situations:
\begin{enumerate}
    \item The problem solved by the analogue simulation is \emph{proven} to be strictly harder than any problem that is simulable by a classical computer. An example (under plausible complexity theoretic conjectures) is boson sampling \citep{Aaronson2005bs}. 

    \item There is no hardness proof, but the \emph{best known classical algorithms} are not able to solve the problem efficiently. Moreover, the quantum simulator \emph{can scale-up} to to large problem sizes without sacrificing accuracy.  
    An example is Lloyd`s digital quantum simulator \citep{lloyd_universal_1996}.
	
	\item There is no hardness proof and the \emph{best known classical algorithms} are not able to solve the problem efficiently. However, it is \emph{unknown} if the quantum simulator can scale-up to to arbitrary problem sizes without sacrificing accuracy.  
	An example is the experiment by \citet{Higgs:2012} (see Sec.~\ref{higgs cs}). Here, classical computational methods (e.g.\ quantum Monte Carlo) can be used only to simulate certain very restricted parameter regimes. For the full simulation a quantum device is required. 

	\item There are \emph{efficient classical algorithms}, but the \emph{scaling of resources} is more favourable in the quantum setting. An example is the setting recently discovered by \citet{bravyi_quantum_2017}. 

\end{enumerate}

That the target model falls into one of these four classes is a clear pragmatic norm for the practice of analogue quantum simulation.  That is, we require some form of quantum computational advantage based upon one of these scenarios. 

\section{Closing Remarks}
Together we expect that our epistemic and pragmatic norms will support the successful pursuit of analogue quantum simulation and emulation in both present day and future scientific practice. That is, we hope that we have offered advice to scientists that they themselves will find useful on their own terms. Our principal goal has been to clarify and analyse ideas that are to a great extent already implicitly endorsed by the scientific community. As such, the normative dimension of our analysis is as much about encouraging scientists to engage in greater explicit reflective methodological discourse, as it is in directing such a discourse in a particular direction. That said, we believe the distinction between analogue quantum simulation and analogue quantum emulation to be an important and timely one. We expect that its adoption would be of lasting benefit to the scientific community.        

\section*{Acknowledgements}
We are extremely grateful to Radin Dardashti, Pete Evans, Sam Fletcher and Hugh Reynolds for helpful comments on a draft manuscript. 
DH thanks Ulrich Schneider and Immanuel Bloch for enlightening in-depth conversations about their perspective on the epistemic goals of cold-atom quantum simulators.
Furthermore, we are grateful to audiences at the annual conference of the BSPS in Edinburgh in July and ECAP9 in Munich in August 2017 for comments on earlier versions of this work. 
DH is supported by the Templeton foundation.
JC is supported by the European Union's Horizon 2020 research and innovation programme under the Marie Sklodowska-Curie grant agreement No.\ 751016. KT is supported by an Arts and Humanities Research Council (UK) grant No.\ AH/P004415/1.

\bibliography{LitAnalogueSim,jc_bib}


\end{document}